\documentclass[11pt]{article}
\usepackage{amssymb,amsmath,amsthm}	% Mathematische Symbole
\usepackage{a4wide}
\usepackage[utf8x]{luainputenc}               	% Eingabe von Umlauten und ß
\usepackage{graphicx}      										% Grafiken einbinden
\usepackage{cite}
\usepackage{color}
\usepackage[numbers]{natbib}
\usepackage{dsfont}
\usepackage{hyperref}
\usepackage{multirow}
\usepackage{subcaption}
\usepackage{amsmath}
\usepackage{xspace}
\usepackage{cleveref}

\newcommand{\R}{\ensuremath{\mathbb{R}}}

\newcommand{\N}{\ensuremath{\mathbb{N}}}

\begin{document}

	%-- TITEL ----------------------------------------------------------%

	\title{Novel Insights in the Levy-Levy-Solomon Agent-Based Economic Market Model}

	\author{ Maximilian Beikirch\footnote{RWTH Aachen University, Templergraben 55, 52056 Aachen, Germany}~\footnote{ORCiD IDs: Maximilian Beikirch: 0000-0001-6055-4089,Torsten Trimborn:  0000-0001-5134-7643},  Torsten Trimborn\footnote{IGPM, RWTH Aachen, Templergraben 55, 52056 Aachen, Germany}~\footnotemark[2]~\footnote{Corresponding author: trimborn@igpm.rwth-aachen.de}}

\maketitle

   %-- INHALTSVERZEICHNIS ----------------------------------------------------------%

%	\tableofcontents

	%-- EINLEITUNG ----------------------------------------------------------%

\begin{abstract}
The Levy-Levy-Solomon model \cite{levy1994microscopic} is one of the most influential agent-based economic market models. In several publications this model has been
discussed and analyzed. Especially Lux and Zschischang \citep{zschischang2001some} have shown that the model exhibits finite-size effects.  
In this study we extend existing work in several directions. First, we show simulations which reveal finite-size effects of the model.
Secondly, we shed light on the origin of these finite-size effects. Furthermore, we demonstrate the sensitivity of the Levy-Levy-Solomon model with respect to random numbers.
Especially, we can conclude that a low-quality pseudo random number generator has a huge impact on the simulation results. 
Finally, we study the impact of the stopping criteria in the market clearance mechanism of the Levy-Levy-Solomon model.
 \end{abstract}

\section{Introduction}
 The Levy-Levy-Solomon model has been introduced by the same-named authors in 1994 \cite{levy1994microscopic}. This model can be regarded as an early contribution in the field of econophysics. 
 In several subsequent publications the authors Levy, Levy and Solomon have extesively studied their model \cite{levy1995microscopic, levy1996complex, levy2000microscopic}. In addition, the model has been
 investigated by other scientists \cite{zschischang2001some, kohl1997influence}. As emphasized by several authors, the Levy-Levy-Salomon model is regarded as one of the most influential agent-based economic market models \cite{samanidou2007agent}. \\ \\
 The model considers heterogeneous financial agents equipped with a personal wealth and their investment decision. The agents have to decide whether they want to invest their money in a safe bond or in risky stocks. The stock price is determined by the market clearance condition which ensures that supply matches demand. The dynamics is driven by a stochastic dividend process. By a utility maximization process, the investment propensity of agents is determined.\\ \\
 The model behavior is studied by means of Monte Carlo simulations which is a classical tool from statistical physics.  
 The original goal of the Levy-Levy-Solomon model was to build a simple model able to reproduce realistic stock price data. In order to judge the quality of artificial stock price data it is common practice to compare statistical quantities of financial data (so-called \textit{stylized facts}), to the artificial data. Examples are the wealth inequality obtained by Pareto \cite{pareto1897cours} or volatility clustering discovered by Mandelbrot \cite{mandelbrot1997variation}. For an introduction to stylized facts we refer to \cite{lux2008stochastic, cont2001empirical, cramer2019stylized}.
 The common goal of agent based economic market models, and to some extend also the Levy-Levy-Solomon model, is to find sufficient conditions for the appearance of stylized facts. 
 In fact, Levy, Levy and Solomon first focused on the reproducibility of chaotic prices and booms and crashes \cite{levy1994microscopic, levy1995microscopic}. In a subsequent publication \cite{levy2000microscopic}, Levy, Levy and Solomon claimed that their model was able to reproduce stylized facts such as volatility clustering and fat-tails in asset returns. \\ \\
Further studies \cite{zschischang2001some, kohl1997influence} have raised doubt on the claimed ability of the Levy-Levy-Solomon model to reproduce stylized facts. 
More precisely it has been shown that the Levy-Levy-Solomon model exhibits finite-size effects, meaning that the output (and hence the ability to reproduce stylized facts) heavily depends on the number of agents in the model. This is an undesireable model characteristic and has been discussed in the context of agent based financial market models in \cite{egenter1999finite, hellthaler1996influence}.\\ \\
In this work, we study the reasons for the appearance of finite size effects in the Levy-Levy-Solomon model and we conduct additional novel studies.
It has been documented in \cite{beikirch2018simulation} that the model behavior of the Levy-Levy-Solomon model is very sensitive to the noise level of the dividend process. We show that a low quality random number generator leads to different qualitative simulation results. Furthermore, we study the impact of different stopping criteria of the market clearance mechanism with respect to the investment decision of agents and several stylized facts. All simulations have been conducted with the recently introduced SABCEMM tool \cite{trimborn2018sabcemm}. \\ \\
 This paper is structured as follows: In the next section we give a detailed definition of the Levy-Levy-Solomon model. In section 3, we give a short introduction to the SABCEMM simulator. 
 In the subsequent section, we present our simulation results. First, we present finite-size effects of the model and discuss their origin. Secondly, we show the impact of a non-reliable (pseudo) random number generator on the qualitative output of the model. Finally, we discuss the impact of different stopping criteria in the clearance mechanism of the model. We finish this work with a short conclusion. 

\section{Levy-Levy-Solomon Model}\label{model}
In this section, we briefly define the Levy-Levy-Solomon (LLS) model. For detailed information regarding the modeling and motivation we refer to the original papers \cite{levy1994microscopic, levy1995microscopic}. The time step $\Delta t>0$ is normalized to one and the time discretization is defined as $t_k=k\cdot 1 ,\ k\in \mathbb{N}$.
The model considers $N\in\N$ financial agents who can invest $\gamma_i\in [0.01, 0.99],\ i=1,...,N$ of their wealth $w_i\in\R_{>0}$ in stocks and have to invest $1-\gamma_i$ of their wealth in a safe bond with interest rate $r\in(0,1)$. The investment propensities $\gamma_i$ are determined by a utility maximization and the wealth dynamic of each agent at time $t\in[0,\infty )$ is given by

\begin{footnotesize}
\begin{align*}
&{w}_i(t_k)=w_i(t_{k-1}) + \left((1-\gamma_i(t_{k-1}))\ r\ w_i(t_{k-1})+\gamma_i(t_{k-1})\ w_i(t_{k-1})\ \underbrace{\frac{{S}(t_k)-S(t_{k-1})+D(t_k)}{S(t_{k-1})}}_{=:x(S,t_k,D)}\right).
\end{align*} 
\end{footnotesize}

The dynamics is driven by a multiplicative dividend process given by:
\[
D(t_k):=(1+ \tilde{z})\  D(t_{k-1}),
\]
 where $\tilde{z}$ is a uniformly distributed random variable with support $[z_1,z_2]$.  
The price is fixed by the so-called \textit{market clearance condition}, where $n\in\N$ is the fixed number of stocks and $n_i(t)$ the number of stocks of each agent.  
\begin{align}
n=\sum\limits_{i=1}^N n_i(t_k)=\sum\limits_{k=1}^N \frac{\gamma_k(t_k)\ w_k(t_k)}{S(t_k)}. \label{fixedpointLLS}
\end{align}

The utility maximization is given by
\[
\max\limits_{\gamma_i \in [0.01,0.99]} E[\log(w(t_{k+1},\gamma_i,S^h))].
\]
where
\begin{align*}
E[\log(w(t_{k+1}, \gamma_i,S^h))]=\frac{1}{m_i} \sum\limits_{j=1}^{m_i} U_i\Bigg(&(1-\gamma_{i}(t_k)) w_{i}(t,S^h) \left(1+r \right)\\ 
&+\gamma_{i}(t_k) w_{i}(t_k, S^h) \Big(1+x\big(S,t_{k-j},D\big) \Big)\Bigg).
\end{align*}
The constant $m_i$ denotes the number of time steps each agent looks back. Thus, the number of time steps $m_i$ and the length of the time step $\Delta t$ defines the time period each agent extrapolates the past values. The superscript $h$ indicates that the stock price is uncertain and needs to be fixed by the market clearance condition. In practice the agents derive their investment proportions $\gamma_i(t_k)$ according to the utility maximization which depends on the hypothetical stock price $S^h$. Provided the market clearance is satisfied the stock price gets fixed, otherwise the hypothetical stock price gets updated and the procedure repeats. Finally, the computed optimal investment proportion gets blurred by a noise term.
\[
\gamma_i (t_k)=H(\gamma_i^{*}(t_k)+\epsilon_i),
\]
where $\epsilon_i$ is a Gaussian distributed random variable with mean zero and standard deviation $\sigma_{\gamma}$.
Here, $H$ denotes the cutoff function which ensures that $\gamma_i\in [0.01,0.99]$ holds. After the noising process, the price is updated. Since the investment fraction is constant we are able to compute the stock price explicitly:
$$
S(t_k)=\frac{\frac{1}{n} \sum\limits_{i=1}^N\gamma_i(t_k) \Big(  w_i(t_{k-1})+  w_i(t_{k-1})\big(  \gamma_i(t_{k-1})\frac{D(t_{k-1})-S(t_{k-1})} { \ S(t_{k-1})   } +(1-\gamma_i(t_{k-1}))\ r  \big) \Big)}{1-\frac{1}{n } \sum\limits_{i=1}^N  \frac{\gamma_i(t_k)\gamma_i(t_{k-1}) w_i(t_{k-1})}{S(t_{k-1})}}.
$$

\paragraph{Utility maximization}
Thanks to the simple log utility function and linear dynamics we can compute the optimal investment proportion in the cases where the maximum is reached at the boundaries. In these cases, the solution is found after two evaluations of $f$, i.e. in constant time.
The first order necessary condition is given by:
$$
f(\gamma_i):= \frac{d}{dt} E[\log(w(t_{k+1}, \gamma_i,S^h))] = \frac{1}{m_i} \sum\limits_{j=1}^{m_i} \frac{ (x\big(S,t_{k-j},D\big)-r)}{(x\big(S,t_{k-j},D\big)-r)\ \gamma_i+ 1+ r}.
$$
 Thus, for $f(0.01)<0$ we can conclude that $\gamma_i=0.01$ holds. In the same manner, we get $\gamma_i=0.99$, if $f(0.01)>0$ and $f(0.99)>0$ holds. 
 Hence, solutions in the interior of $[0.01, 0.99]$ can be only expected in the case: $f(0.01)>0$ and  $f(0.99)<0$. This coincides with the observations in \cite{samanidou2007agent}.

\section{The SABCEMM Simulator}\label{sabcemm}
\begin{figure}[h!]
\begin{center}
\includegraphics[width=0.45\textwidth]{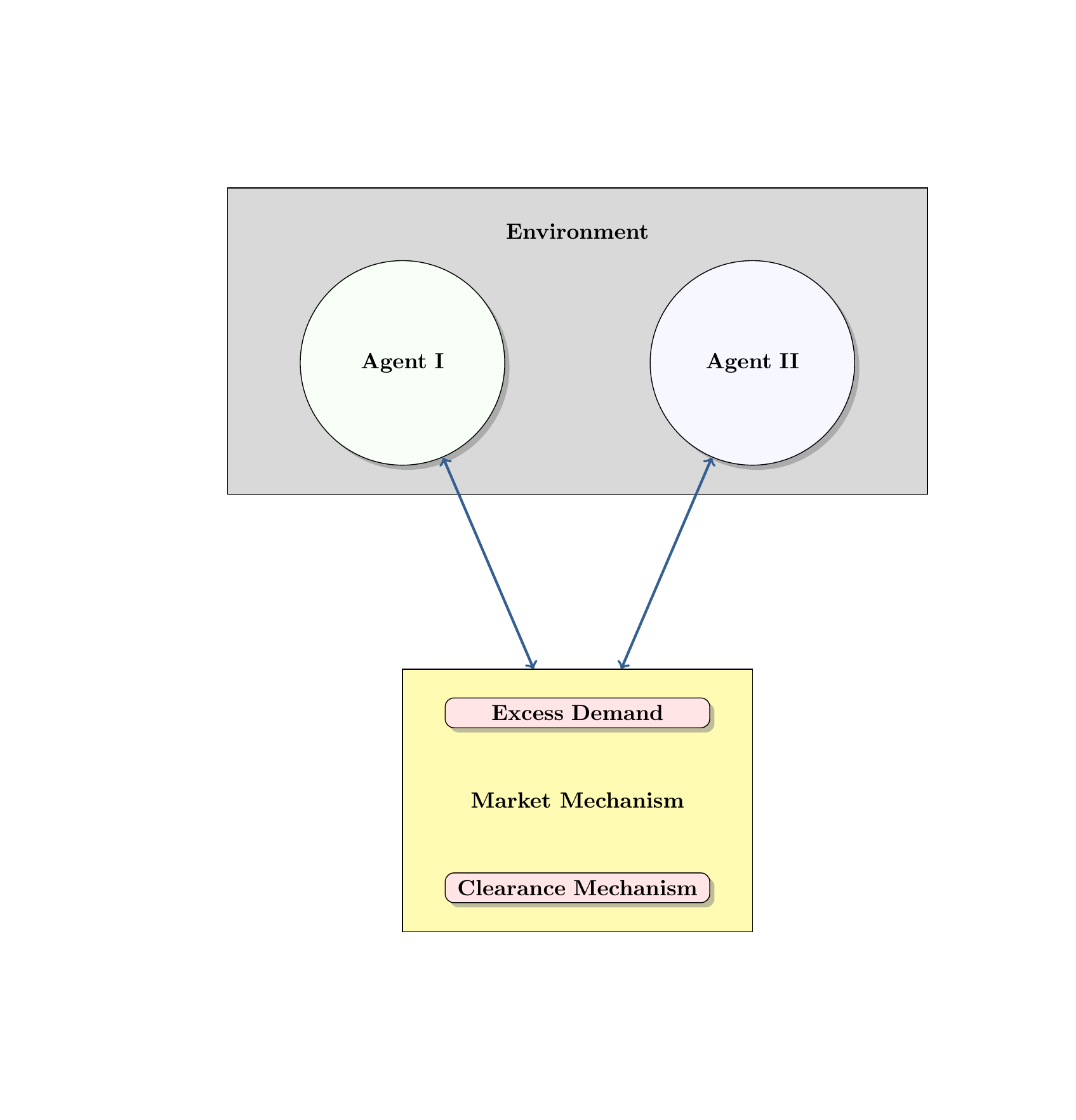}
\caption{Schematic picture of the abstract ABCEMM model \citep{trimborn2018sabcemm}. }\label{OurModel}
\end{center}
\end{figure}

The simulation results in this study have been created by the  recently introduced open source simulator SABCEMM \citep{trimborn2018sabcemm}.
SABCEMM  is especially designed for large-scale simulations of agent based computational economic market (ABCEM) models, which is essential for the study of finite-size effects.
This simulator implements an object oriented design leveraging a generalized structure of ABCEM models as defined in \citep{trimborn2018sabcemm}.
The implementations of the individual ABCEM building blocks are well-separated and the ABCEM model is assembled from the building blocks via an XML-based configuration file.
Hence, the evaluation of an ABCEM model using a different building block, such as the market mechanism, requires only a change in the configuration file.
If the changed building block does not already exist, only this single block has to be implemented.
In the following, we present the main conceptual ideas behind the simulator.
\\ \\
SABCEMM is well suited for any economic market model which consists of at least one \textit{agent} and one \textit{market mechanism}.
An agent is an investor who has a supply of or demand for a certain good or asset, which is traded at the market. 
The market mechanism determines the price from the demand and supply of all market participants.
More precisely, we differentiate between the so-called \textit{price adjustment process} and the \textit{excess demand calculator}.
The latter one aggregates the supply and demand of all market participants (agents) to a single quantity, the excess demand.
The former one represents the method of how the market price is fixed based on this excess demand.
A schematic picture, which illustrates the presented ideas is shown in Figure \ref{OurModel}.
The concept of an \textit{environment} has been first introduced by the authors in \citep{trimborn2018sabcemm}.
An environment represents possible additional coupling between the agents.
Probably, the most famous example for an environment is herding, which is frequently used in ABCEM models.
We emphasize that such an environment is not mandatory.
For a rigorous mathematical definition of the meta-model which is the foundation of the SABCEMM simulator and a detailed discussion of technical details and computational aspects of SABCEMM, we refer to \citep{trimborn2018sabcemm}. We finish the presentation of the SABCEMM simulator by an short introduction of the supported pseudo random number generators.  \\ \\

In SABCEMM, we provide two pseudo random number generators (RNG) based on the widely popular Mersenne Twister family of RNGs.
First, we include the \texttt{MT19937} (64-bit) Mersenne Twister, introduced into the C++ standard library with the C++11 standard.
\texttt{MT19937} provides an extremely long period of $2^{19937}-1 \approx 10^{6001.6}$ and particularly good equidistribution (623 dimensions of approximate equidistribution \cite{PRAND}).
Secondly, we support the \texttt{MT2203} (64-bit) multi-stream Mersenne twister implemented within the Intel Math Kernel Library (MKL)\footnote{\url{https://software.intel.com/en-us/mkl} (may require extra license)} providing a period of $2^{2203}-1 \approx 10^{663.2}$, very good equidistribution (68 dimensions of approximate equidistribution \cite{PRAND}) and 6024 parallel streams.
Note that \texttt{MT2203} is included as it outperforms the \texttt{MT19937} implementation within the C++11 standard library by far when combined with pooling of generated random numbers.
For further details on pooling of random numbers within SABCEMM, the interested reader is refered to \cite{trimborn2018sabcemm}.
Since \texttt{MT19937} and \texttt{MT2203} fail only on two and four tests of the BigCrush battery of statistical tests \cite{TestU01}, we accept these highly popular RNGs as high-quality RNGs.
In comparison, in section \ref{sec:Pseudo Random Numbers}, we show that the low-quality RNG \texttt{RANDU} \cite{RANDU}, failing already 14 tests within the SmallCrush and 125 tests within the Crush battieries \cite{TestU01}, is of insufficient quality for application within ABCEM models.
RNGs within SABCEMM are seeded from the \texttt{/dev/urandom} device on Linux/Unix systems.
If the device is not available, the current time stamp\footnote{\texttt{std::chrono::high\_resolution\_clock::now()}} is used.

\section{Simulation Results}
\label{examples}
In this section, we perform several numerical tests on the LLS Model. First, we show finite-size effects in the LLS model and study their origins. 
Furthermore, we present an example which reveals the sensitivity of the LLS model with respect to (pseudo) random numbers and especially the impact of a low quality pseudo random number generator. 
Finally, we show that the termination condition in the root finding algorithm of the LLS model has a huge impact on the model output.\\ \\
All presented results have been generated by the SABCEMM simulator which is freely available on GitHub \cite{SABCEMMgithub}.
For the pseudo random number generators used in each simulation, please consult Table \ref{tab:RNGTable}.
Furthermore, the simulation data is published \cite{SabnumData} such that the reader can reproduce the presented results.

\subsection{Finite-Size Effects}

\begin{figure}[ht]
\centering
\begin{subfigure}{0.49\linewidth}
\includegraphics[width=\linewidth]{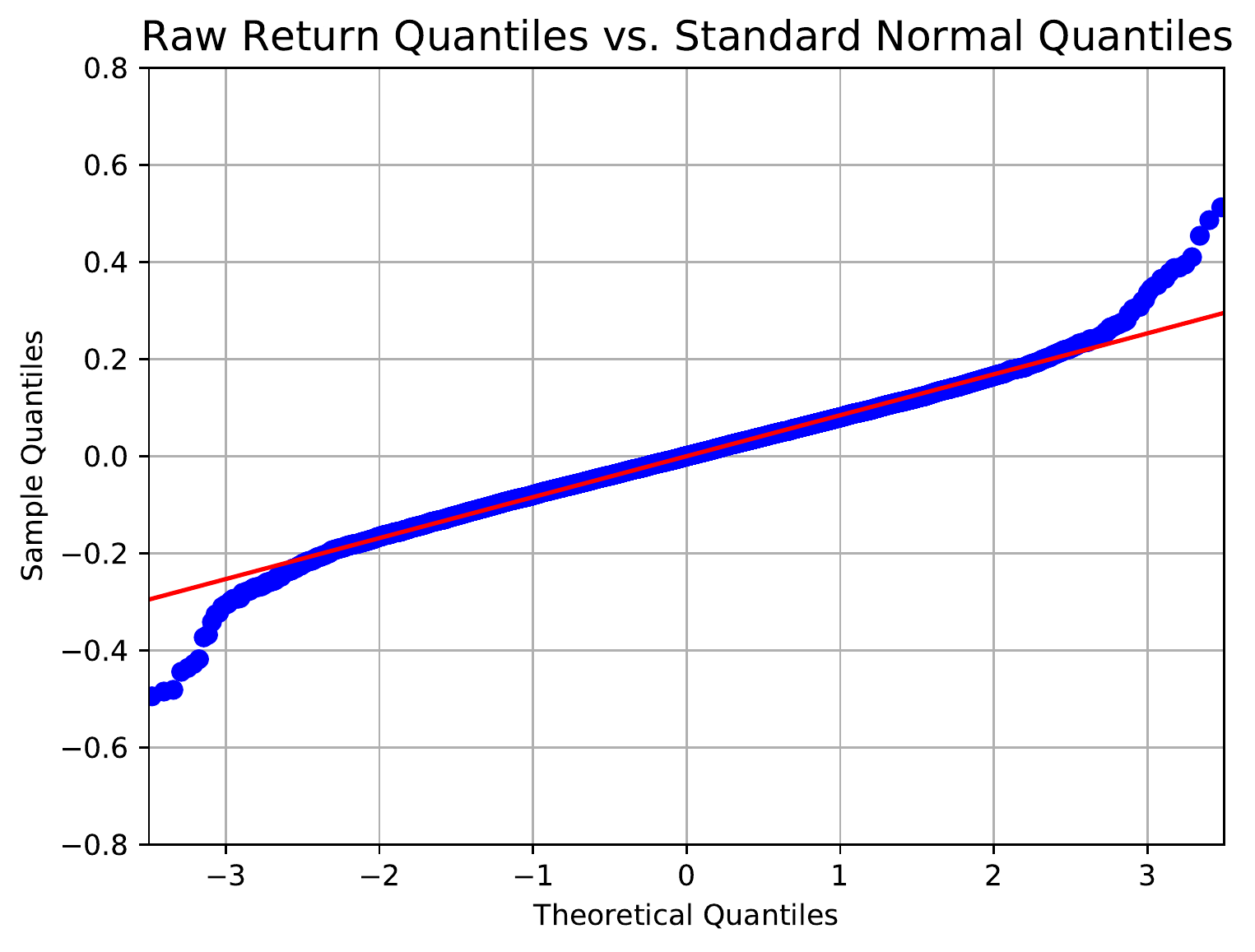}
\caption{QQ-plot of returns with $N=99$. Heavy tails are visible.}
\label{fig:embfinitesize_wealth_a}
\end{subfigure}
\begin{subfigure}{0.49\linewidth}
\includegraphics[width=\linewidth]{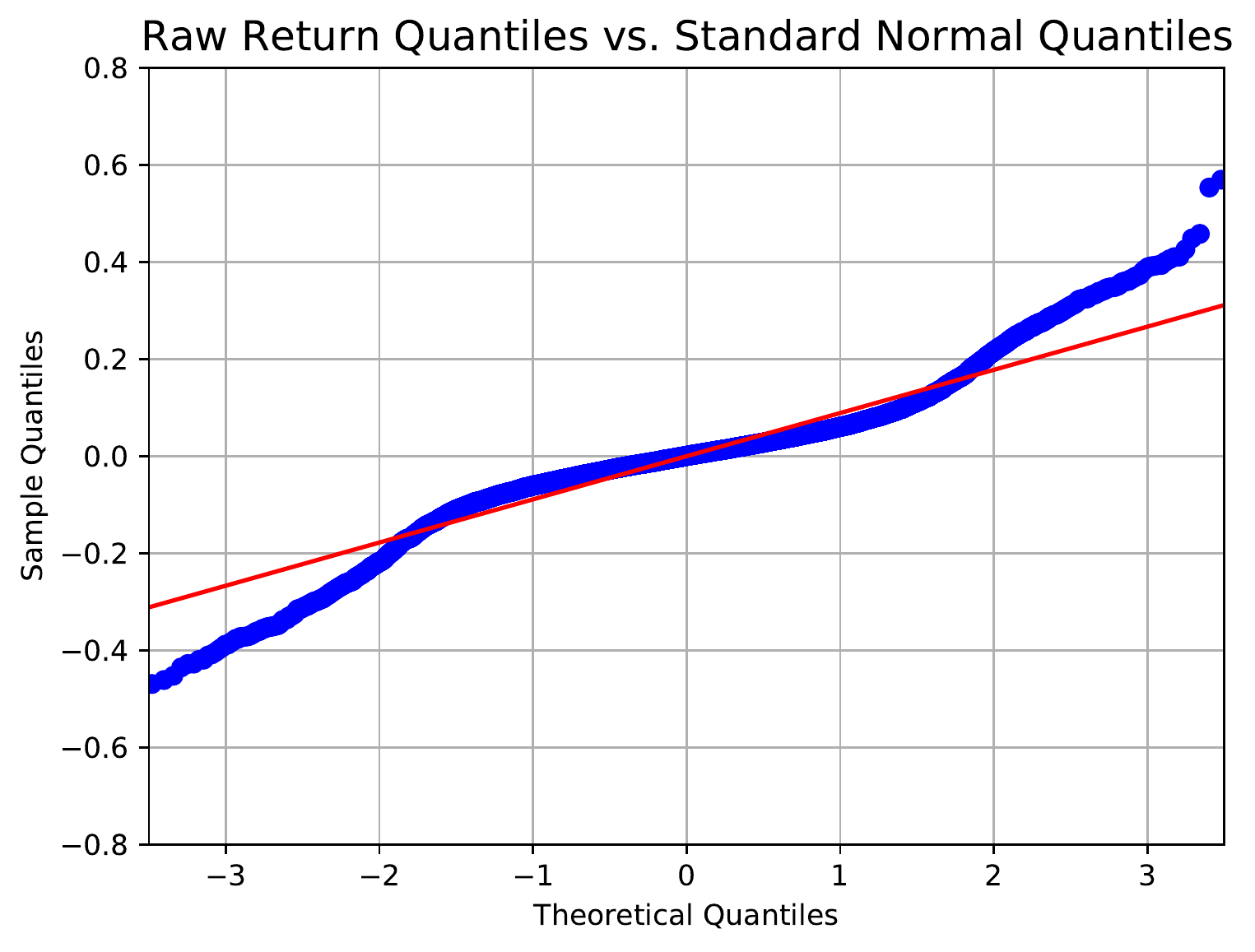}
\caption{QQ-plot of  returns with $N=999$ agents. Heavy tails are visible but changed their shape (cmp. \cref{fig:embfinitesize_wealth_a}).}
\label{fig:embfinitesize_wealth_b}
\end{subfigure}
\begin{subfigure}{0.49\linewidth}
\includegraphics[width=\linewidth]{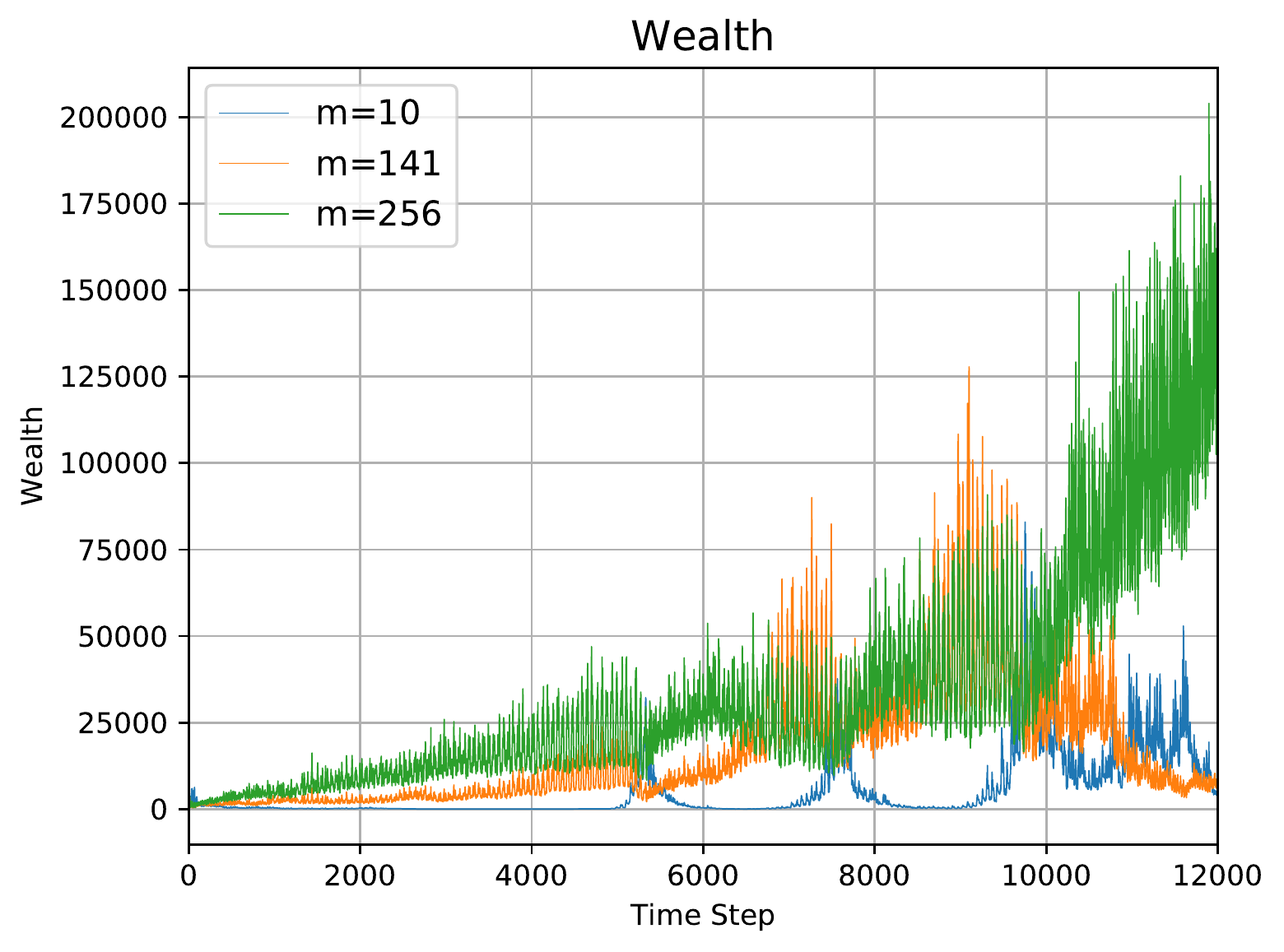}
\caption{Aggregated wealth of each agent group with $N=33$ agents.}
\label{fig:embfinitesize_wealth_c}
\end{subfigure}
\begin{subfigure}{0.49\linewidth}
\includegraphics[width=\linewidth]{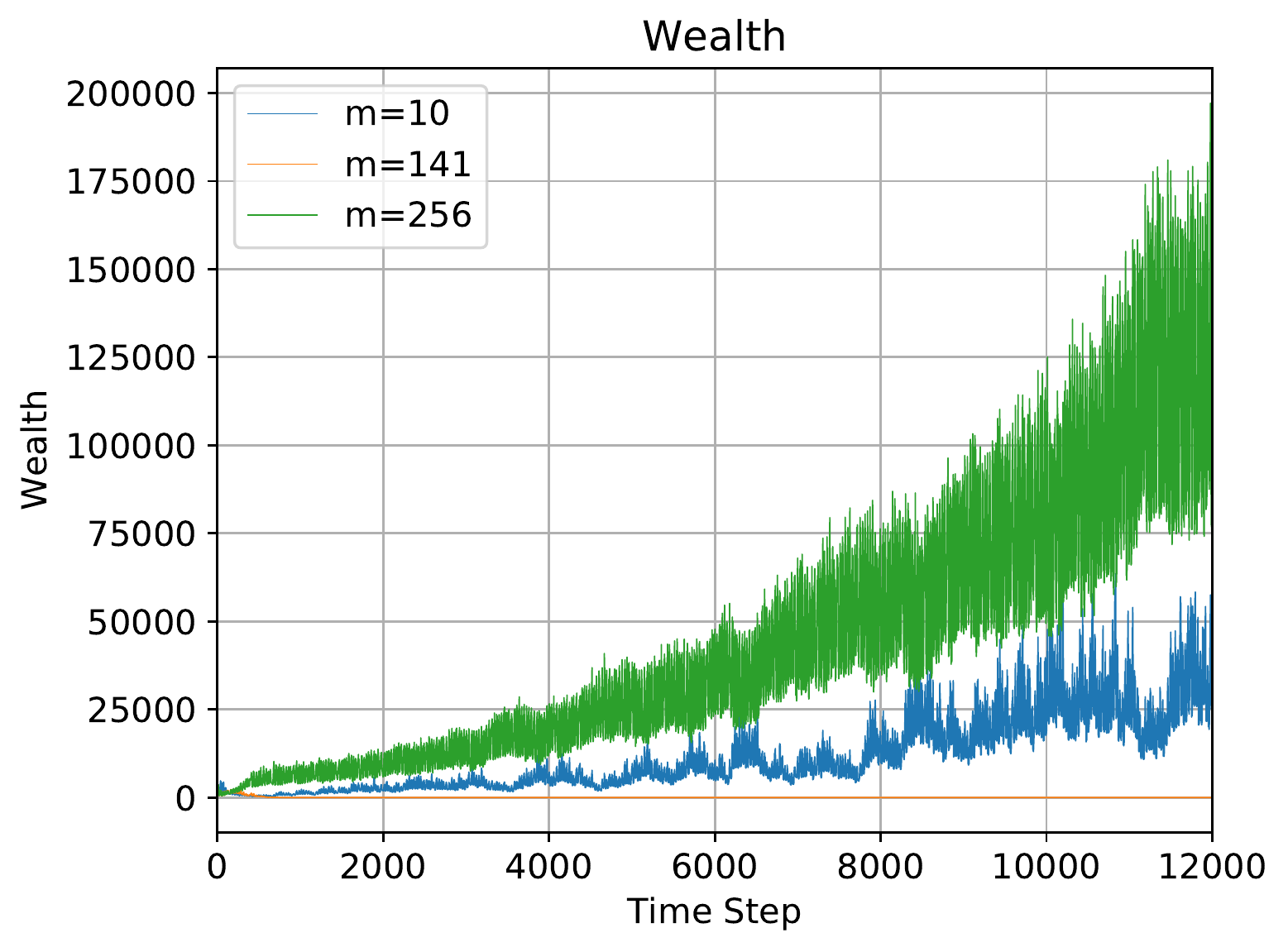}
\caption{Aggregated wealth of each agent group with $N=333$ agents. The group ranking has changed.}
\label{fig:embfinitesize_wealth_d}
\end{subfigure}
\caption{Simulation of the LLS model with 99 agents (left) and 999 agents (right). The agents are divided in three groups with different memory sizes. \Cref{fig:embfinitesize_wealth_a} and \ref{fig:embfinitesize_wealth_b} reveal a change in the tail behavior, while \cref{fig:embfinitesize_wealth_a} and \ref{fig:embfinitesize_wealth_b} shows a change in the group ranking. Parameters as in \cref{LLS-3-agents}. For colored plots, please refer to the online version.}
\label{fig:embfinitesize_wealth}
\end{figure}

Finite-size effects in ABCEM models have been documented by several authors \cite{egenter1999finite, hellthaler1996influence}, in particular for the LLS model \cite{zschischang2001some, kohl1997influence}. 
First, we demonstrate finite-size effects and secondly discover the reasons for that behavior.  \\
\label{MA-LLS}
In our simulations, we observe two different effects caused by varying the number of agents.
First, we obtain that the tail behavior of the wealth distribution changes for an increasing number of agents. Secondly, we find that dependent on the number of agents, different agent types (differing w.r.t. their memory mechanism) dominate the market. A population is considered dominant if it attains maximum wealth. Levy at al. \cite{levy1996complex} claimed that an investor group with higher memory span always dominates a group with a smaller memory.
We have conducted simulations with 99 agents (33 short memory, 33 medium memory, 33 large memory, for details we refer to appendix \ref{parameter}) and with 999 (333 short memory, 333 medium memory, 333 long memory, details in appendix \ref{parameter}) agents, respectively. The aggregated wealth evolution of the three agent groups (Figure \ref{fig:embfinitesize_wealth}) shows that the group ranking changes for different number of agents. 
Furthermore, the quantile-quantile plots in Figure \ref{fig:embfinitesize_wealth} clearly indicate the change in tail behavior for different number of agents. Thus, we can conclude that the qualitative output of the model changes with respect to the number of agents. We highlight that this is an utmost undesirable model characteristic.  

\begin{figure}[htb]
\begin{subfigure}{\linewidth}
\includegraphics[width=0.49\linewidth]{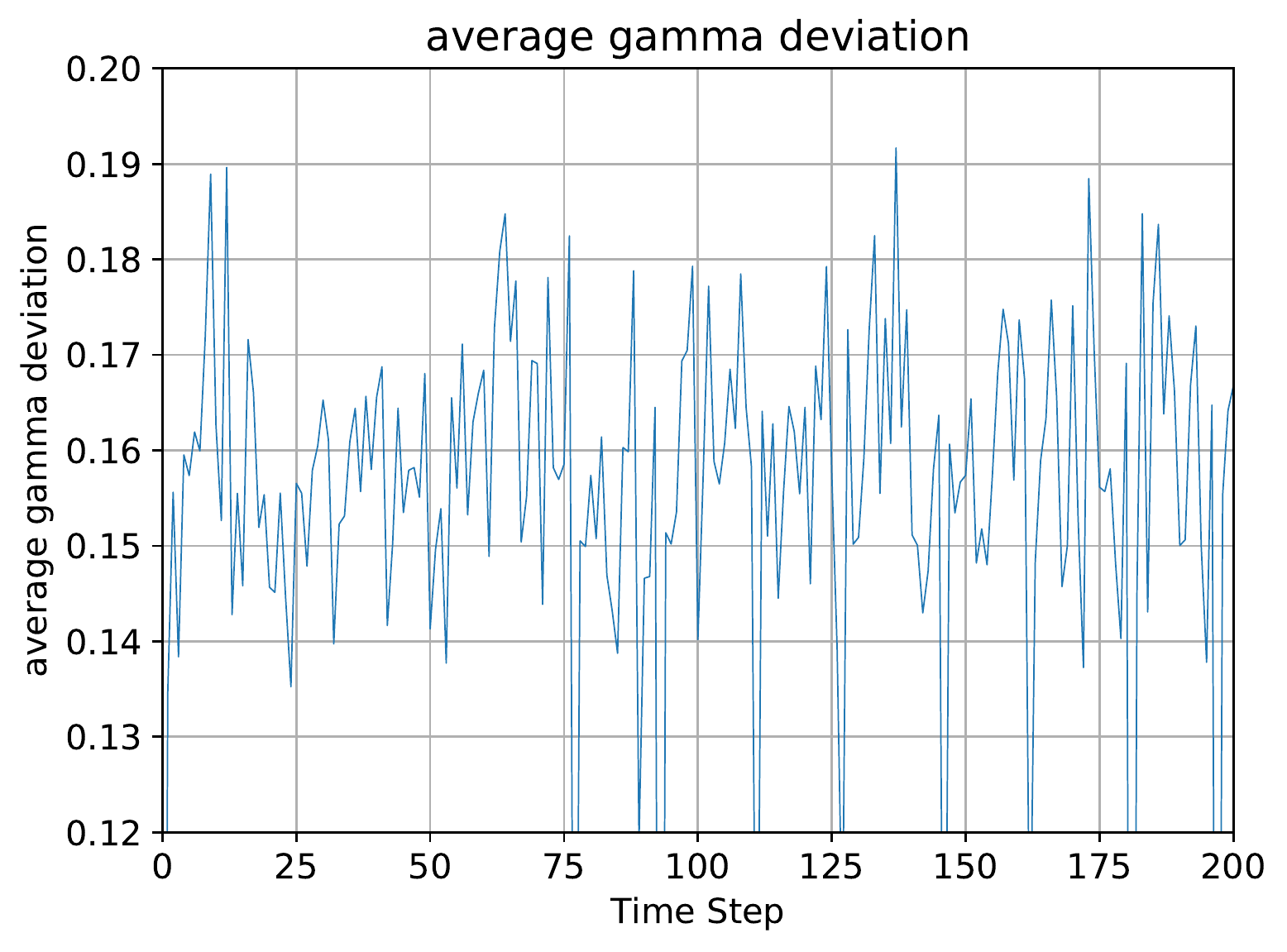}
\includegraphics[width=0.49\linewidth]{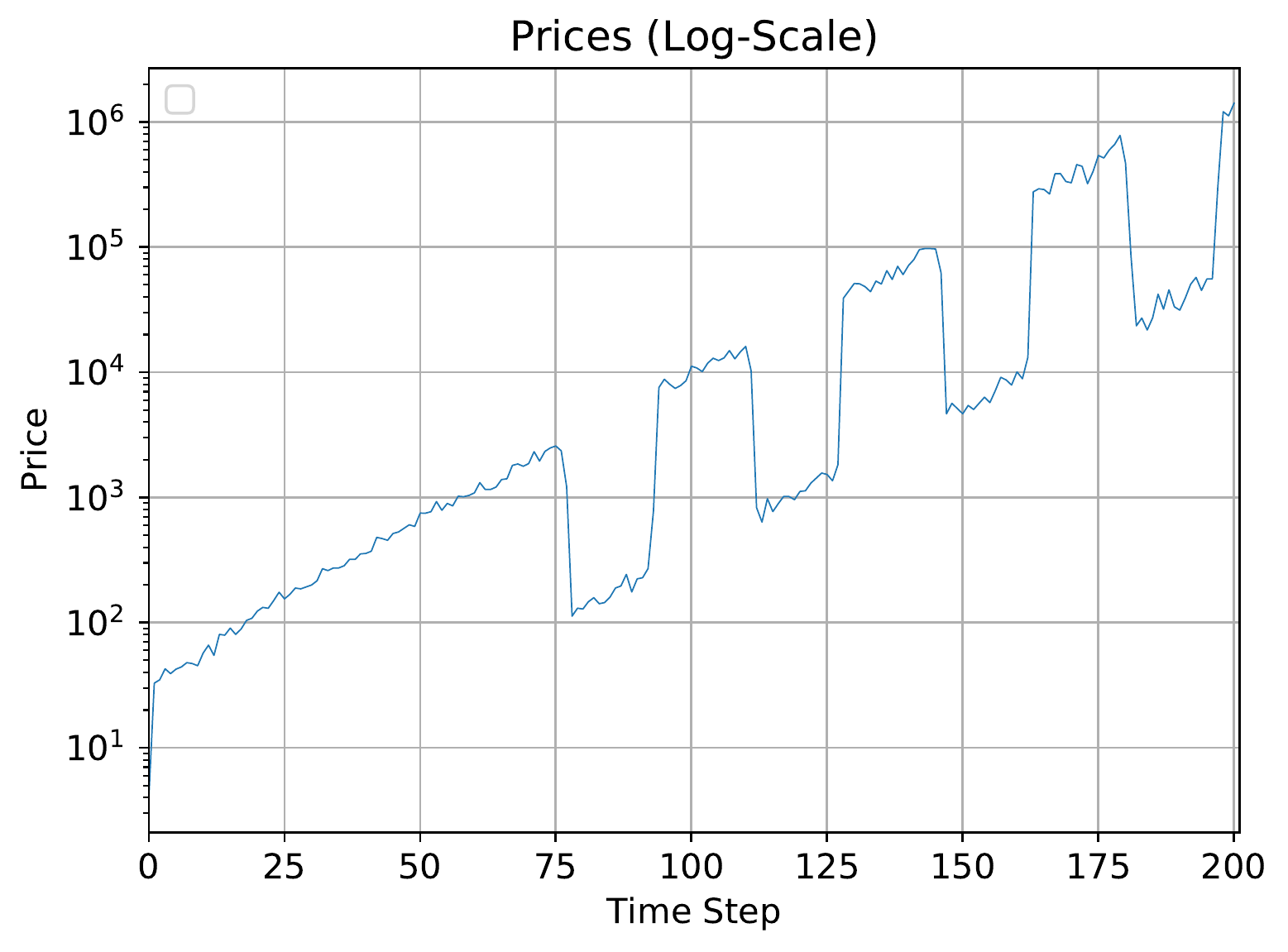}
\caption{200 agents}
\label{fig:emb_fse_erklarung_a}
\end{subfigure}
\begin{subfigure}{\linewidth}
\includegraphics[width=0.49\linewidth]{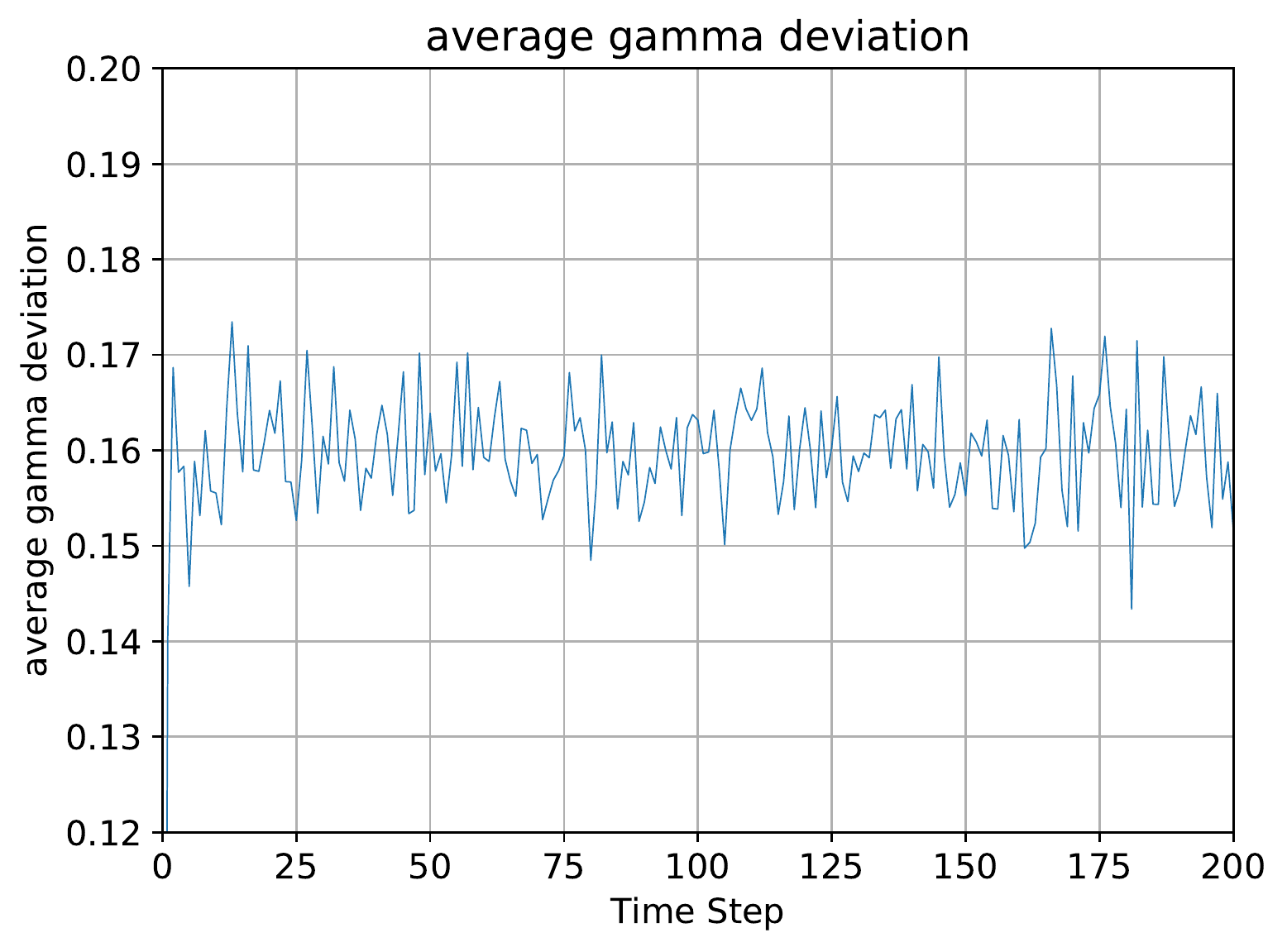}
\includegraphics[width=0.49\linewidth]{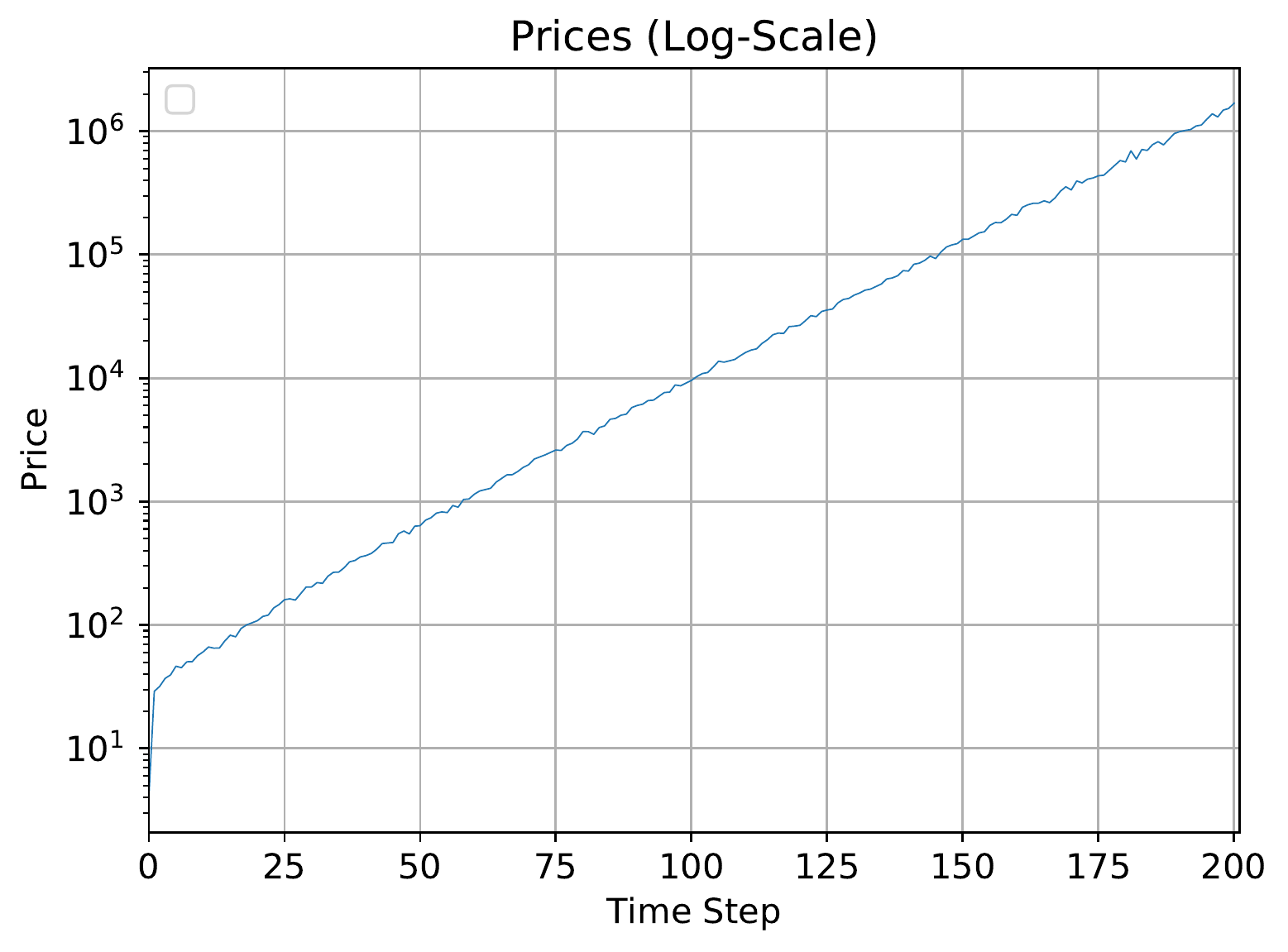}
\caption{500 agents}
\label{fig:emb_fse_erklarung_b}
\end{subfigure}
\begin{subfigure}{\linewidth}
\includegraphics[width=0.49\linewidth]{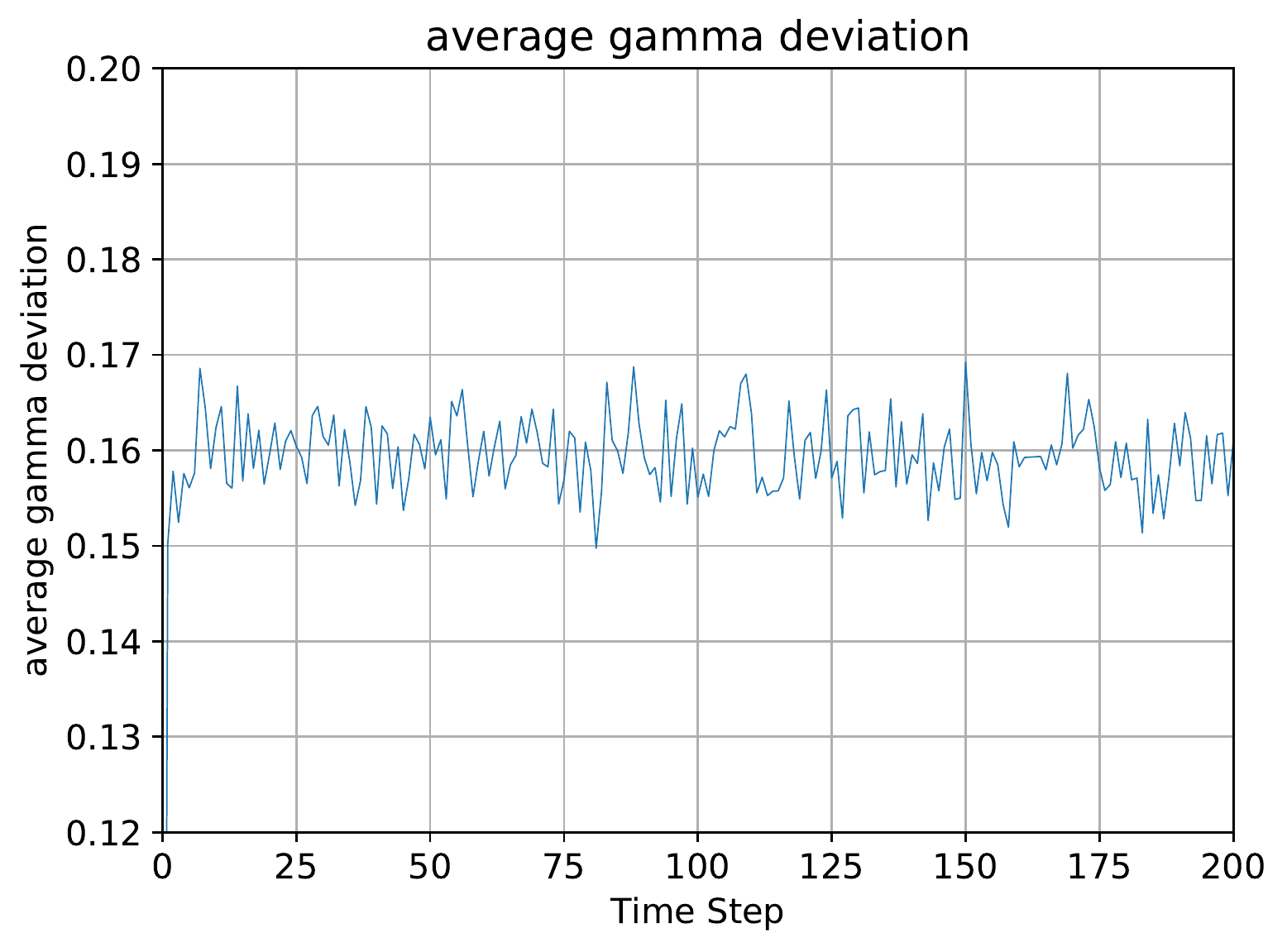}
\includegraphics[width=0.49\linewidth]{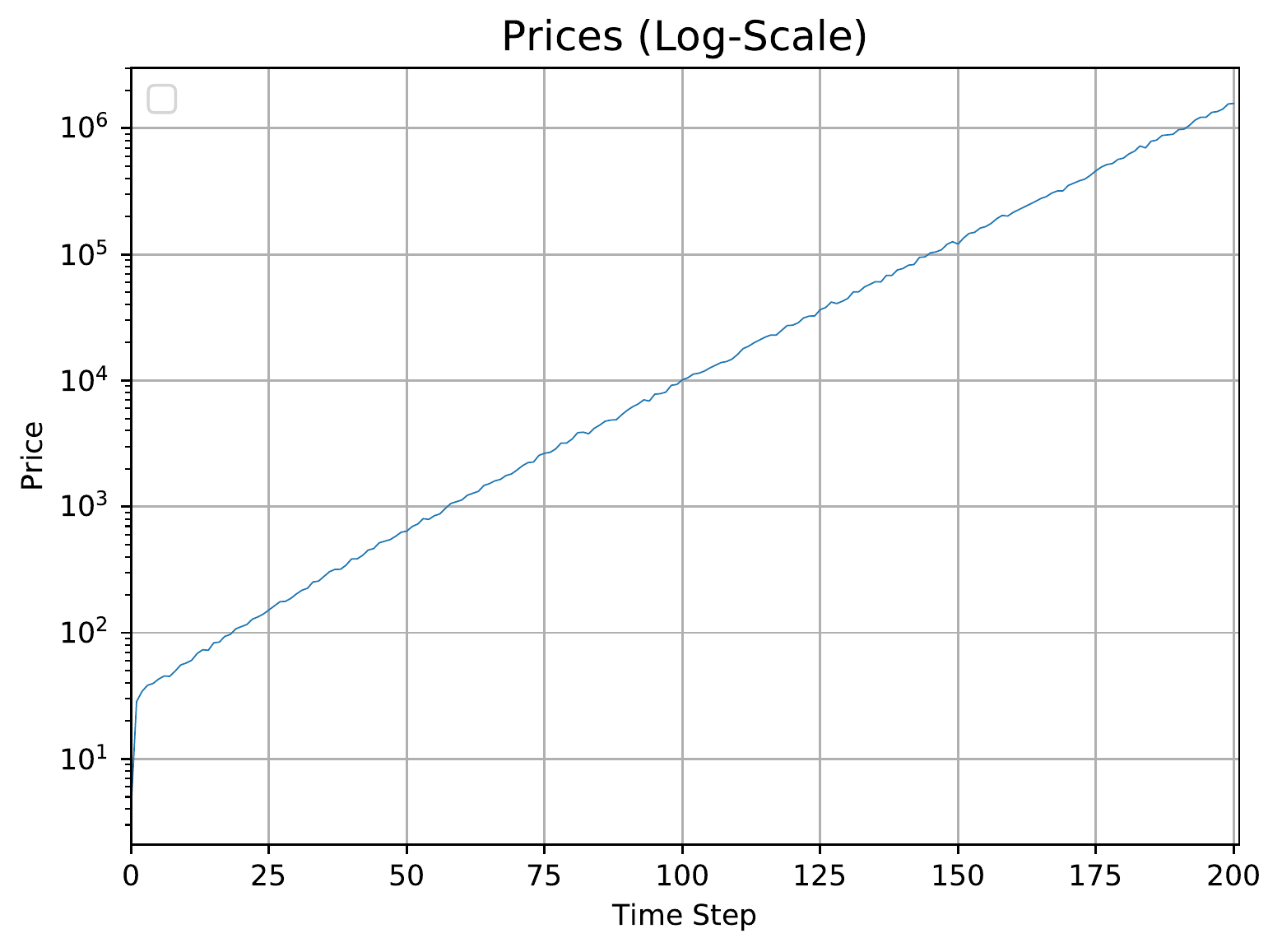}
\caption{1000 agents}
\label{fig:emb_fse_erklarung_c}
\end{subfigure}
\caption{Prices and gamma differences $d_{\gamma}^N$ for different agent counts.}
\label{fig:emb_fse_erklarung}
\end{figure}

We aim to understand why changes of agent count lead to changes in the model behavior.
Before discussing the appearance of finite-size effects, we need to understand the origins of the \textit{usual} oscillatory model behavior.
 Studies by Levy, Levy and Solomon \cite{levy1995microscopic}, Zschischang and Lux \cite{zschischang2001some}, and Otte et al. \cite{beikirch2018simulation} emphasize the importance of white noise on the characteristic oscillatory model behavior. In fact, in the zero noise case the oscillatory model characteristic completely vanishes. 
The reason is that the stock return is always better than the bond return (in the original parameter setting) and thus the investors allocate their money in stocks. This leads to an exponential increase in the price and to constant investment propensities. 
In the case of noise on the investment fraction it is possible that the return on stocks is worse than the return on bonds which may lead to changes in the investment propensity.
These rapid changes in the investment propensity lead to oscillatory price behavior. As shown in \cite{beikirch2018simulation}, the wave period of the wealth evolution heavily depends on the noise level.  Clearly, in the case of no noise, we have no oscillatory behavior since the price is in equilibrium, which is computed by the clearance mechanism:

$$
n=\sum\limits_{i=1}^N n_i(t)= \sum\limits_{i=1}^N \frac{w_i(t)}{S(t)} \gamma_i(t).
$$
In the noisy case one adds to every optimal $\gamma_i^*$ a noise term: $\gamma_i= H(\gamma_i^*+\epsilon_i), $ where $\epsilon$ is a Gaussian random variable.
Then finally one can update the final stock price by an explicit computation since the investment fraction $\gamma_i$ is constant. For details we refer to section \ref{model}. In fact the stock price is computed as a quantity proportional to
$$
S(t)\propto \frac{1}{N} \sum\limits_{i=1}^N  H(\gamma_i^*(t)+ \epsilon_i),
$$
since the number of stocks $n$ scales with the number of agents. In order to quantify the influence of the noise we define the difference of investment fractions before and after noise application:
$$
d_{\gamma}^N(t):=\Big| \frac{1}{N} \sum\limits_{i=1}^N  H(\gamma_i^*(t)+ \epsilon_i)-\frac{1}{N} \sum\limits_{i=1}^N  \gamma_i^*(t)\Big|.
$$
Note that the difference is not additive due to the additional cutoff function $H$. Figure \ref{fig:emb_fse_erklarung} depicts
average difference of investment fractions $d_{\gamma}^N$ and the stock price  for different number of agents. The simulation results clearly indicate that the variance of $d_{\gamma}^N$ decreases for increasing number of agents. Especially, we are able to deduce that large deviations of the mean of the $d_{\gamma}^N$ are needed in order to obtain the typical oscillatory behavior.
Hence, we can conclude that an increasing number of agents reduces the variance of fluctuations caused by additional noising in the LLS model and heavily influences the stock price behavior, resulting in fewer market crashes. Therefore we claim that this scaling behavior with respect to the number of agents causes the finite-size effects in the LLS model.

%\begin{figure}[h!]
%    \begin{center}
%        \includegraphics[width=0.7\textwidth]{figs/LLS/LLSqqLogReturns}
%        \caption{QQ-Plot of log-returns. Simulation conducted with $1,000$ agents and the parameters are set as in \cref{LLS-3-agents} with $\sigma_{\gamma}=0.2$.}
%        \label{OM-LLS-3-4-qq}
%    \end{center}
%\end{figure}
%
%\begin{figure}[h!]
%    \begin{center}
%        \includegraphics[width=0.7\textwidth]{figs/LLS/LLS-Wealth99}
%        \caption{Aggregated wealth of three equally sized agent groups. The total number of agents is $99$, for further parameters we refer to table \ref{LLS-3-agents}. }
%        \label{MA-LLS-cash1}
%    \end{center}
%\end{figure}
%
%\begin{figure}[h!]
%    \begin{center}
%        \includegraphics[width=0.7\textwidth]{figs/LLS/LLS-Wealth999}
%        \caption{Aggregated wealth of three equally sized agent groups. The total number of agents is $999$, for further parameters we refer to table \ref{LLS-3-agents}.}
%        \label{MA-LLS-cash3}
%    \end{center}
%\end{figure}

\clearpage

\subsection{Pseudo Random Numbers}
\label{sec:Pseudo Random Numbers} 
\begin{figure}[tb]
\centering
        \begin{subfigure}{0.35\linewidth}
            \includegraphics[width=\textwidth]{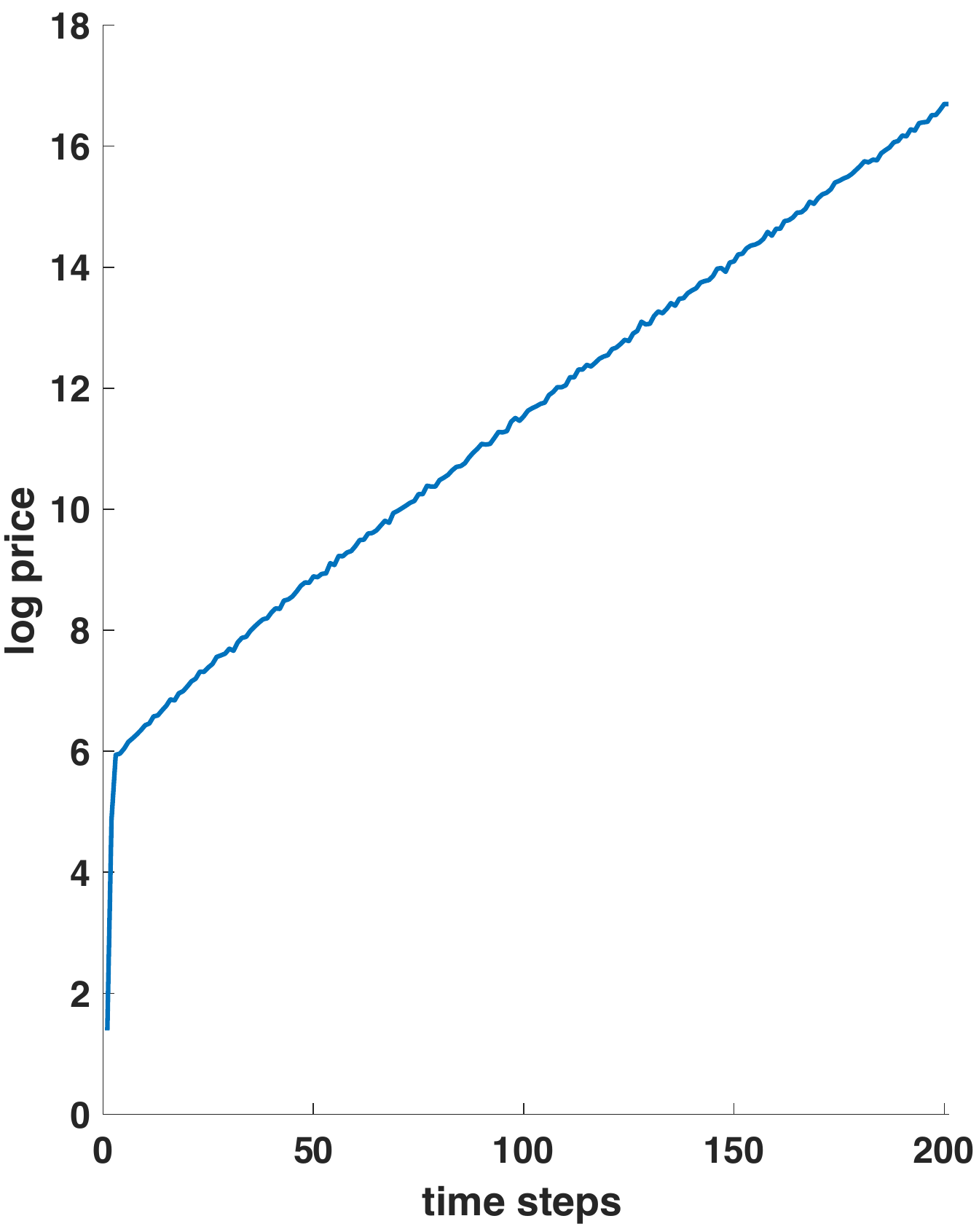}
            \caption{C++ MT19937 RNG ($64$ bit)}
            \label{qual-rng_a}
        \end{subfigure}
        \begin{subfigure}{0.37\linewidth}
    \includegraphics[width=\textwidth]{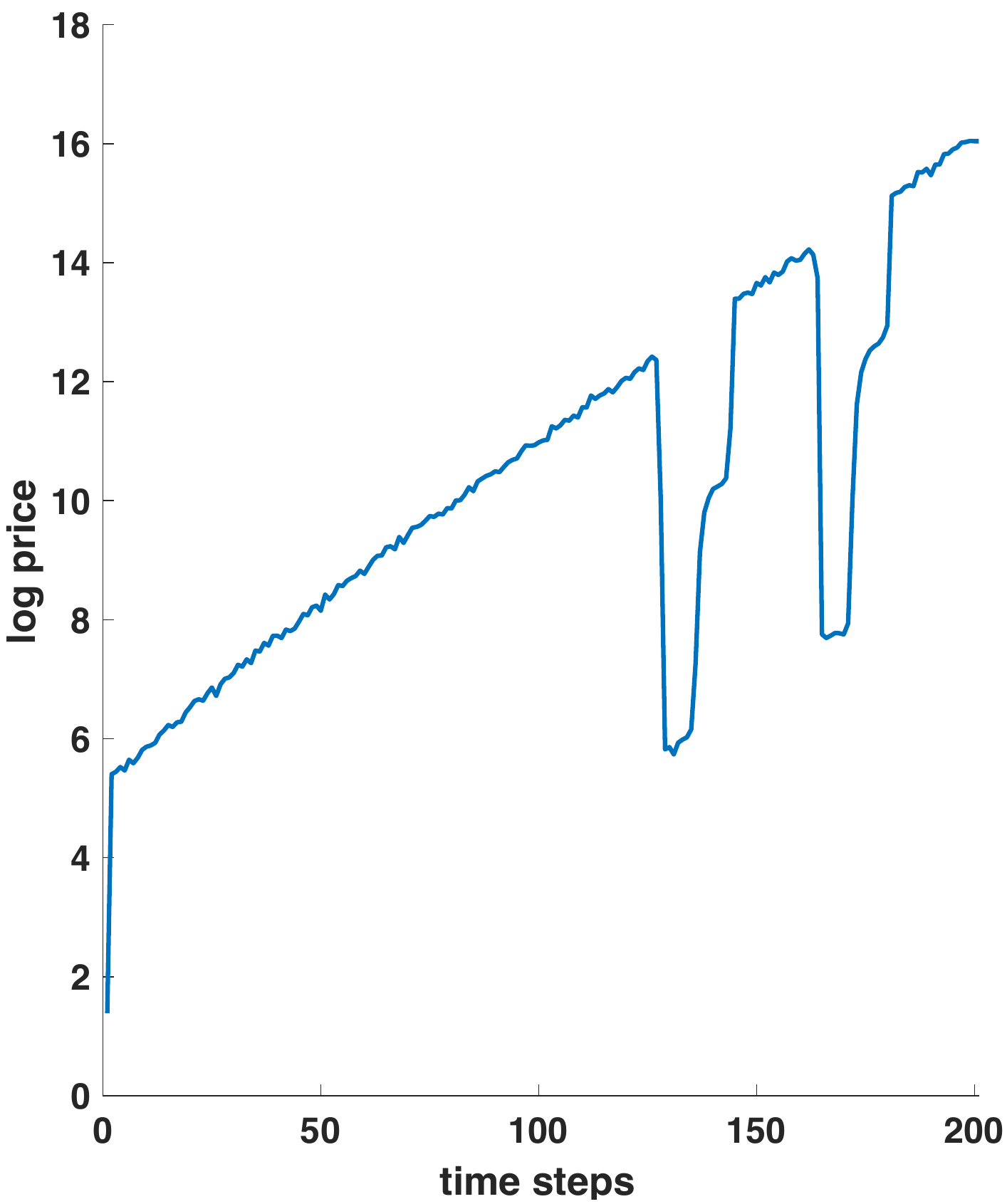}
    \caption{\texttt{RANDU} generator ($32$ bit)}
    \label{qual-rng_b}
\end{subfigure}
        \caption{Simulations of the LLS model conducted with different pseudo random number generators. Parameters as in Table \ref{LLS-basic} with $\sigma_{\gamma}=0.01$.}
        \label{qual-rng}
\end{figure}

As discussed in \cite{trimborn2018sabcemm}, ABCEM models require the generation of huge amounts (commonly in the order of several million) of high-quality pseudo random numbers.
Furthermore, it is well documented that many ABCEM models are sensitive to the precise noise level \cite{beikirch2018simulation}, i.e. the variance of the pseudo random numbers used.
Thus, the question whether low-quality pseudo random numbers influence the qualitative results of ABCEM models is legit.
 We answer this question examplarily by applying the well-known linear congruential pseudo random number generator \texttt{RANDU}, \cite{hellekalek1998good, knuth1997art} to the LLS model. \texttt{RANDU} is known for poor performance when used for the generation of many pseudo random numbers.
We compare an implementation of \texttt{RANDU} on a ARMv7 $32$ bit processor to the C++11 standard pseudo random number generator on a x86\_64 64 bit processor. As shown in Figure \ref{qual-rng}, changing the pseudo random number generator drastically changes the qualitative model output. 
Thus, we may conclude that models sensitive to the choice of random variables, generate different model outputs with respect to different pseudo random number generators. 
Therefore we emphasize that we explicitly do not recommend using the \texttt{RANDU} generator. Due to this, the published version of SABCEMM does not support the \texttt{RANDU} generator as we have only used this generator for this special test.

\clearpage

\subsection{Stopping Criteria in Clearance Mechanism}
\label{NumSolv}
\begin{figure}[ht]
\centering
\begin{subfigure}{0.49\linewidth}
\includegraphics[width=\linewidth]{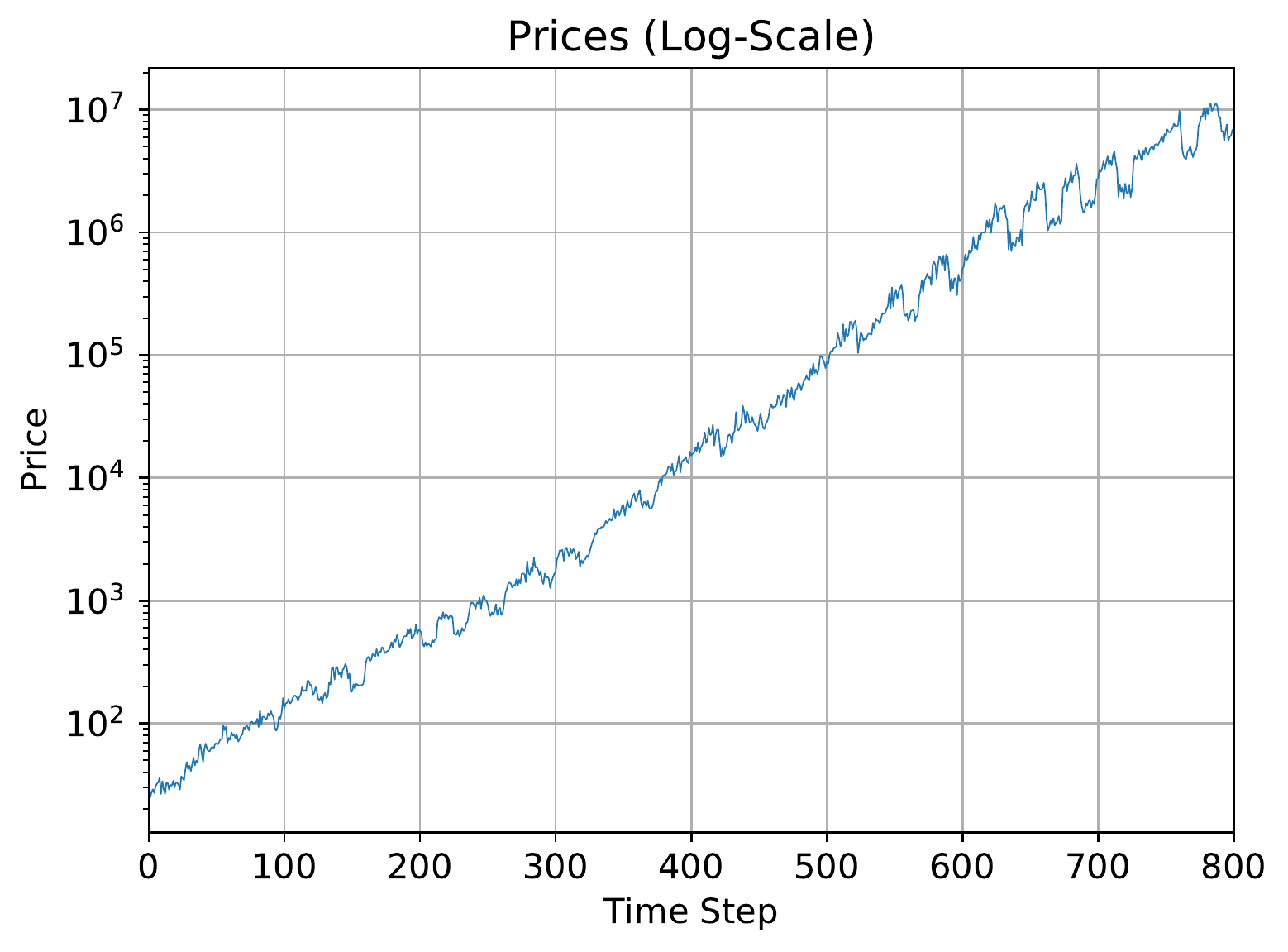}
\includegraphics[width=\linewidth]{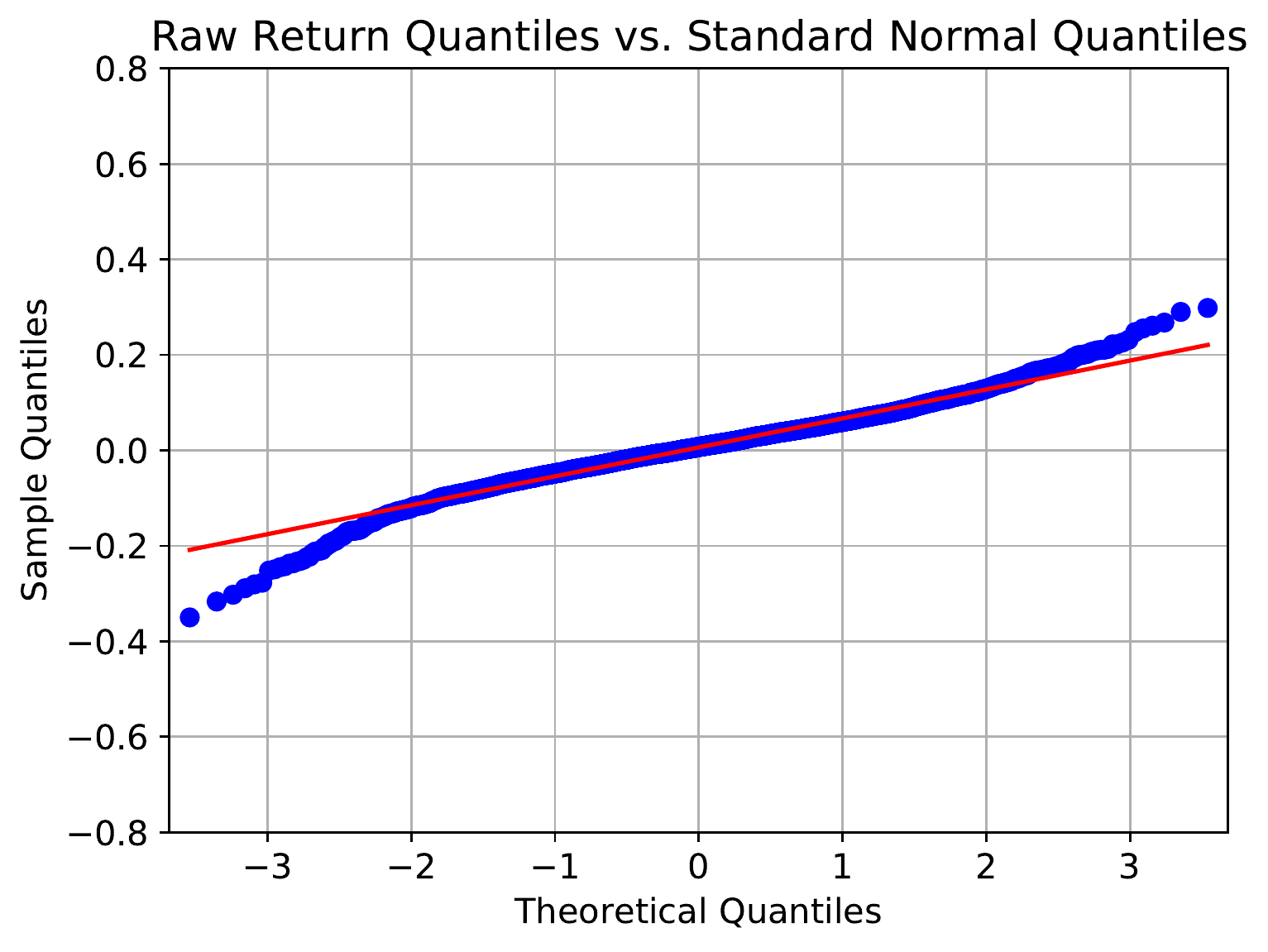}
\caption{$\xi$ = 0.1}
\label{fig:toleranz_vergleich_prices_a}
\end{subfigure}
\begin{subfigure}{0.49\linewidth}
\includegraphics[width=\linewidth]{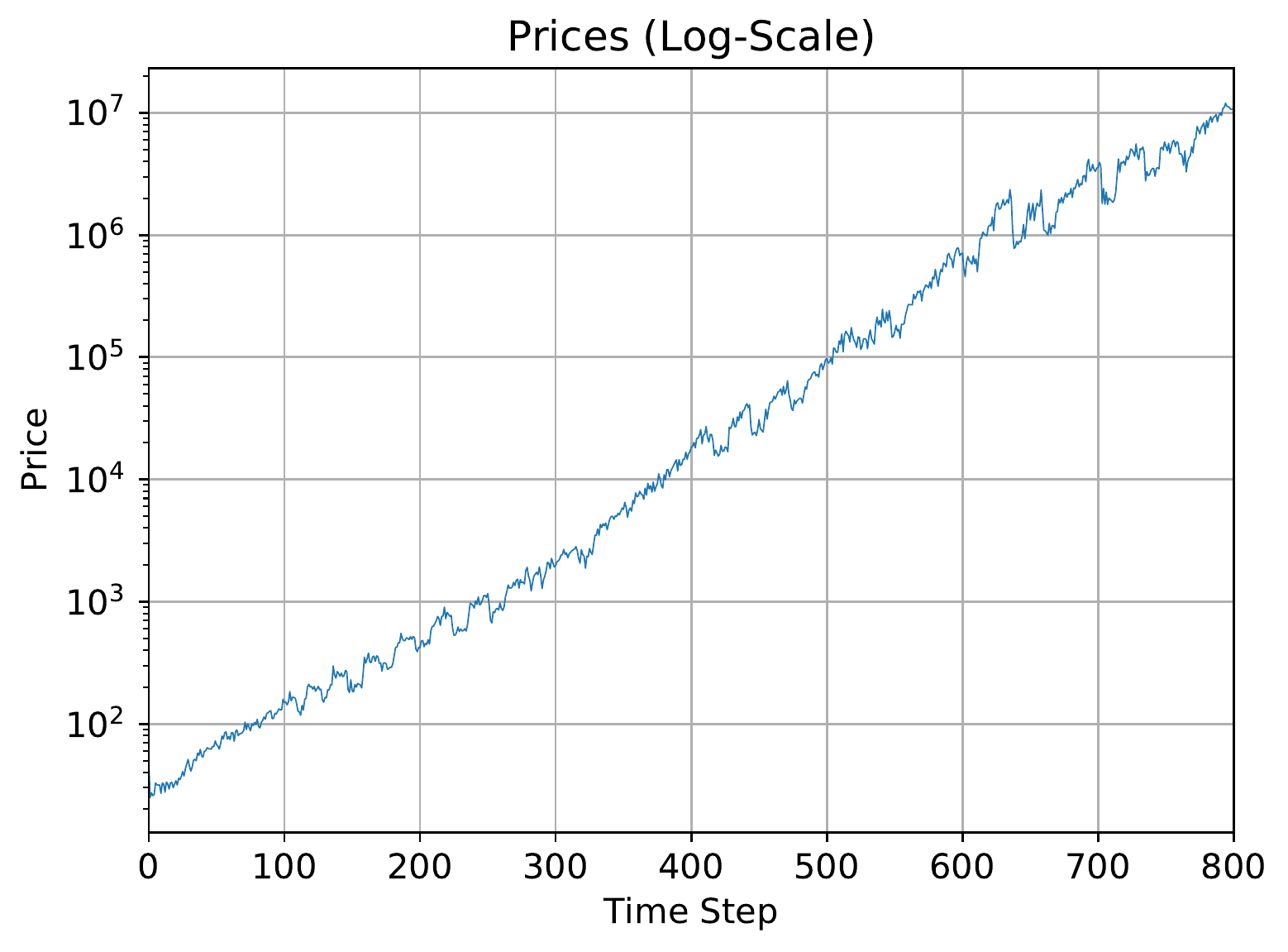}
\includegraphics[width=\linewidth]{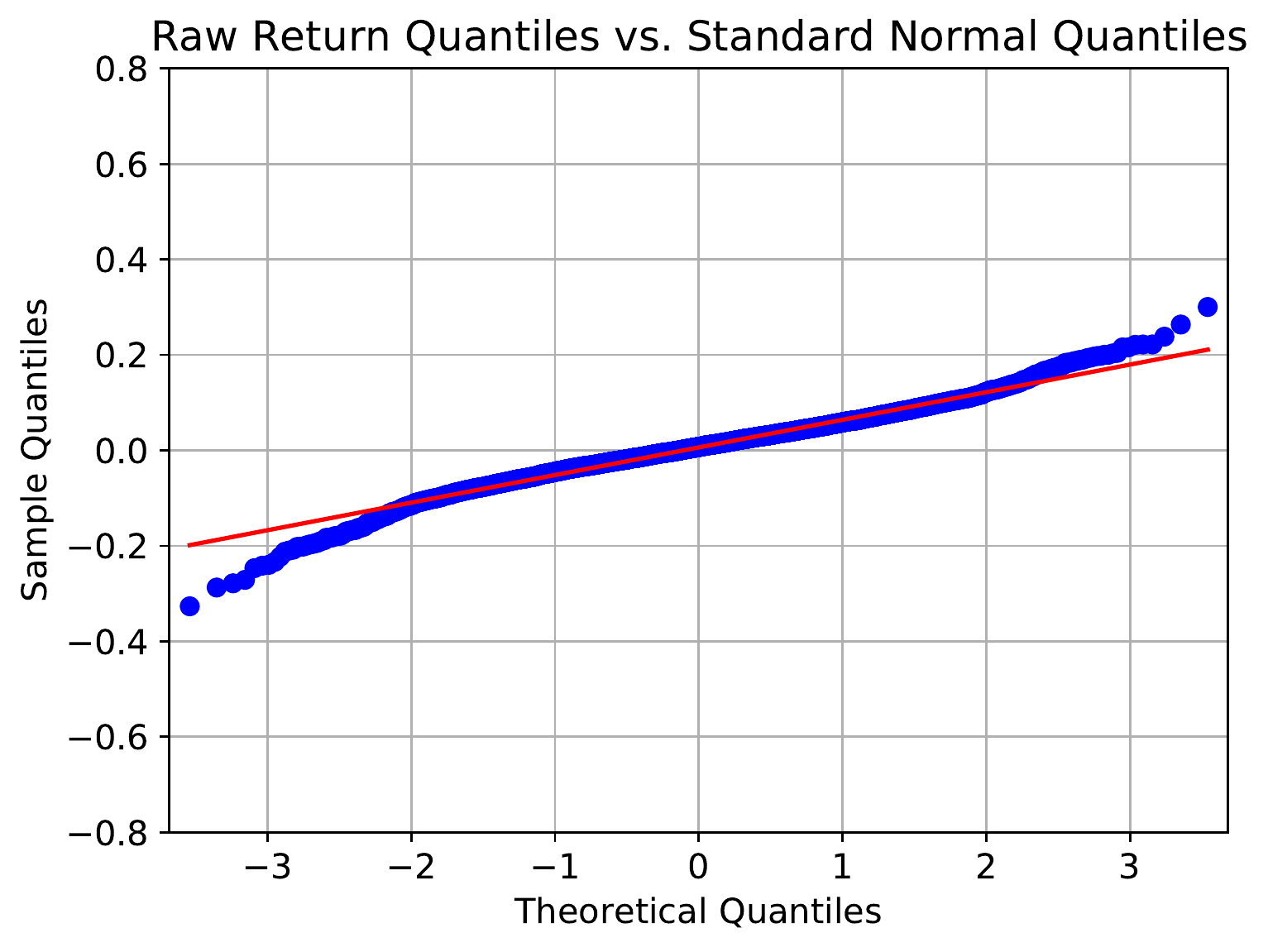}
\caption{$\xi$ = 0.05}
\label{fig:toleranz_vergleich_prices_b}
\end{subfigure}
\caption{Price trends and return distribution for low tolerances.}
\label{fig:toleranz_vergleich_prices}
\end{figure}

\begin{figure}[ht]
\centering
\begin{subfigure}{0.49\linewidth}
\includegraphics[width=\linewidth]{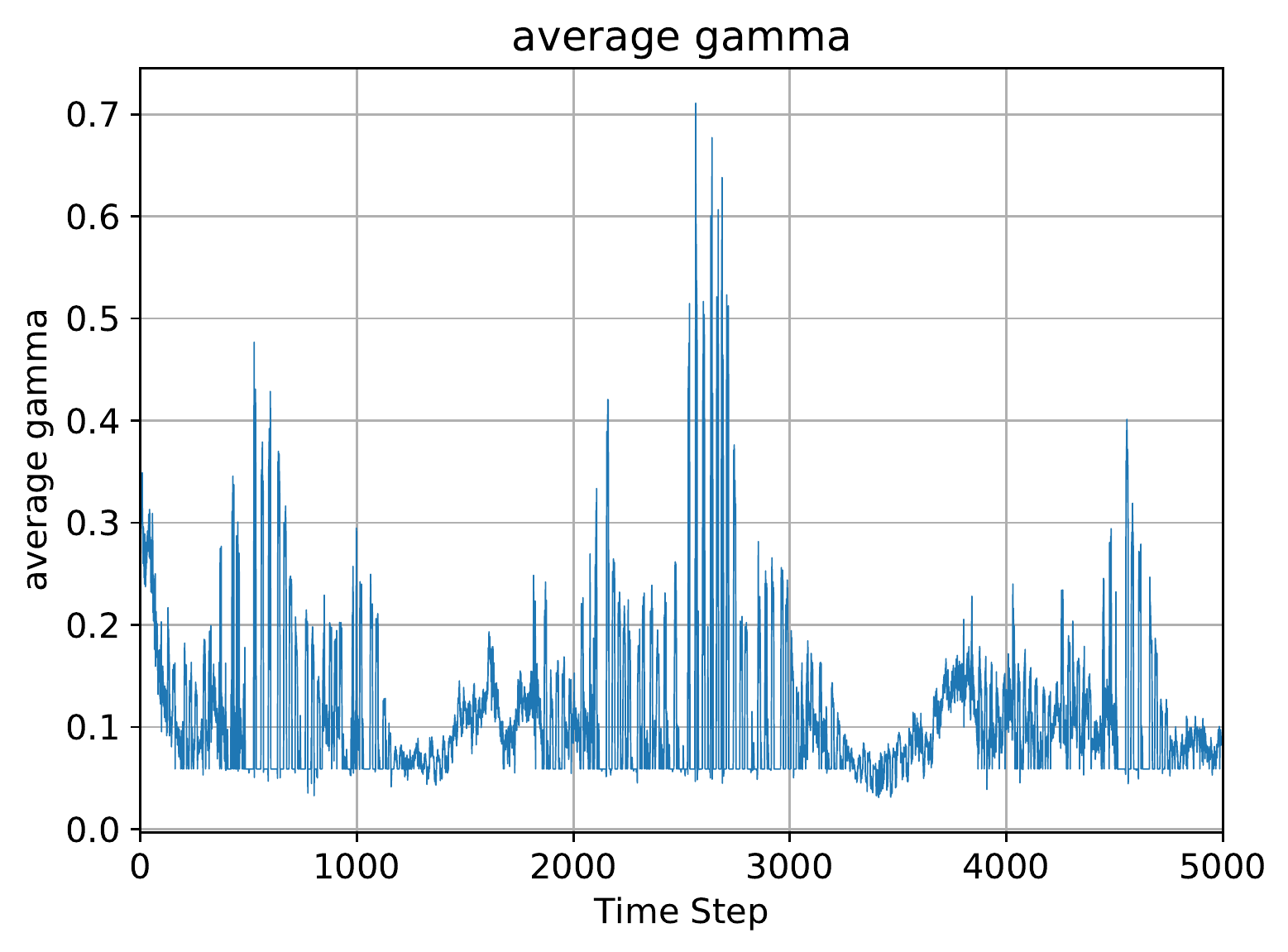}
\caption{$\xi$ = 0.1}
\label{fig:toleranz_vergleich_gamma_a}
\end{subfigure}
\begin{subfigure}{0.49\linewidth}
\includegraphics[width=\linewidth]{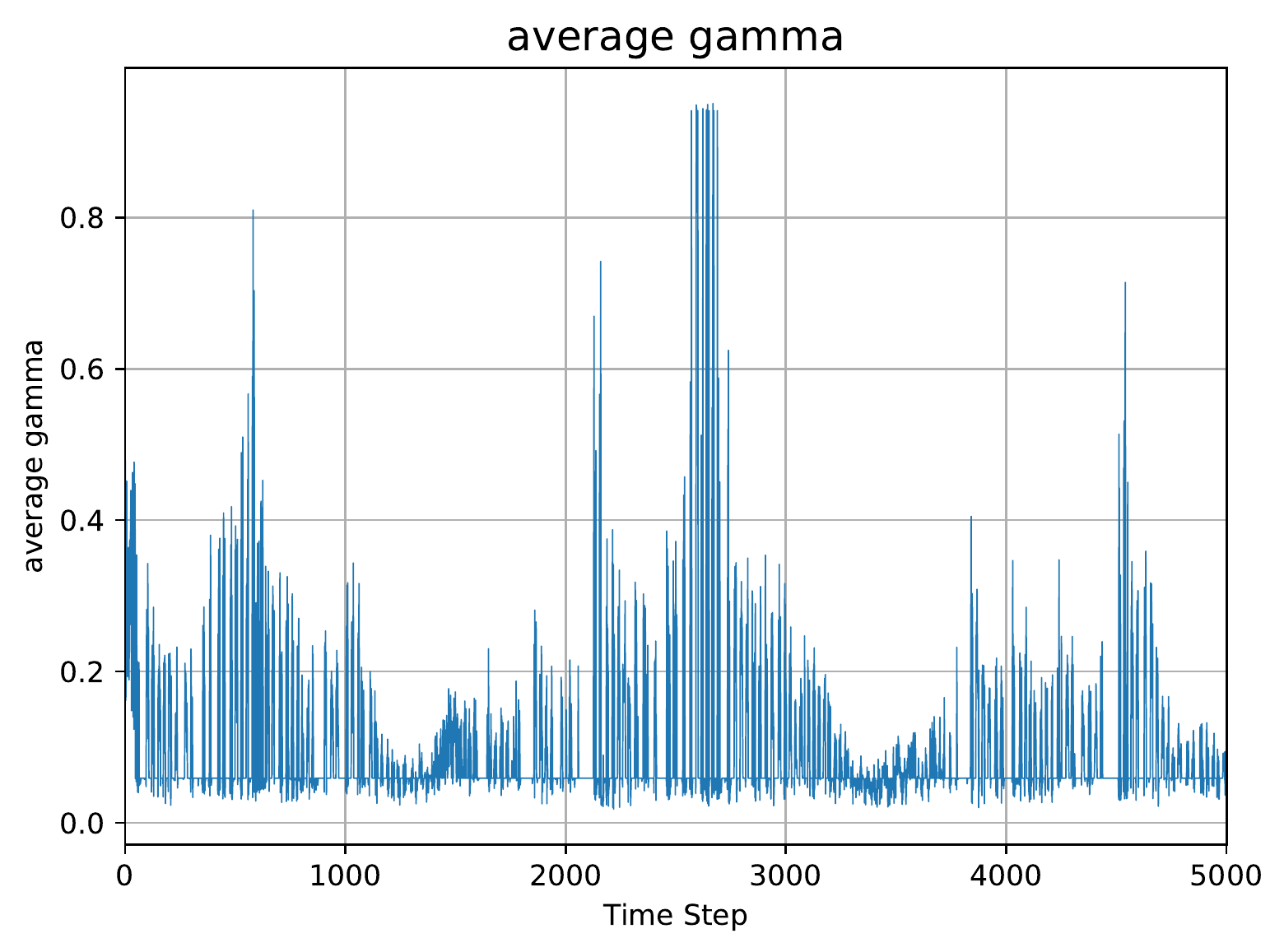}
\caption{$\xi$ = 0.5}
\label{fig:toleranz_vergleich_gamma_b}
\end{subfigure}
\begin{subfigure}{0.49\linewidth}
\includegraphics[width=\linewidth]{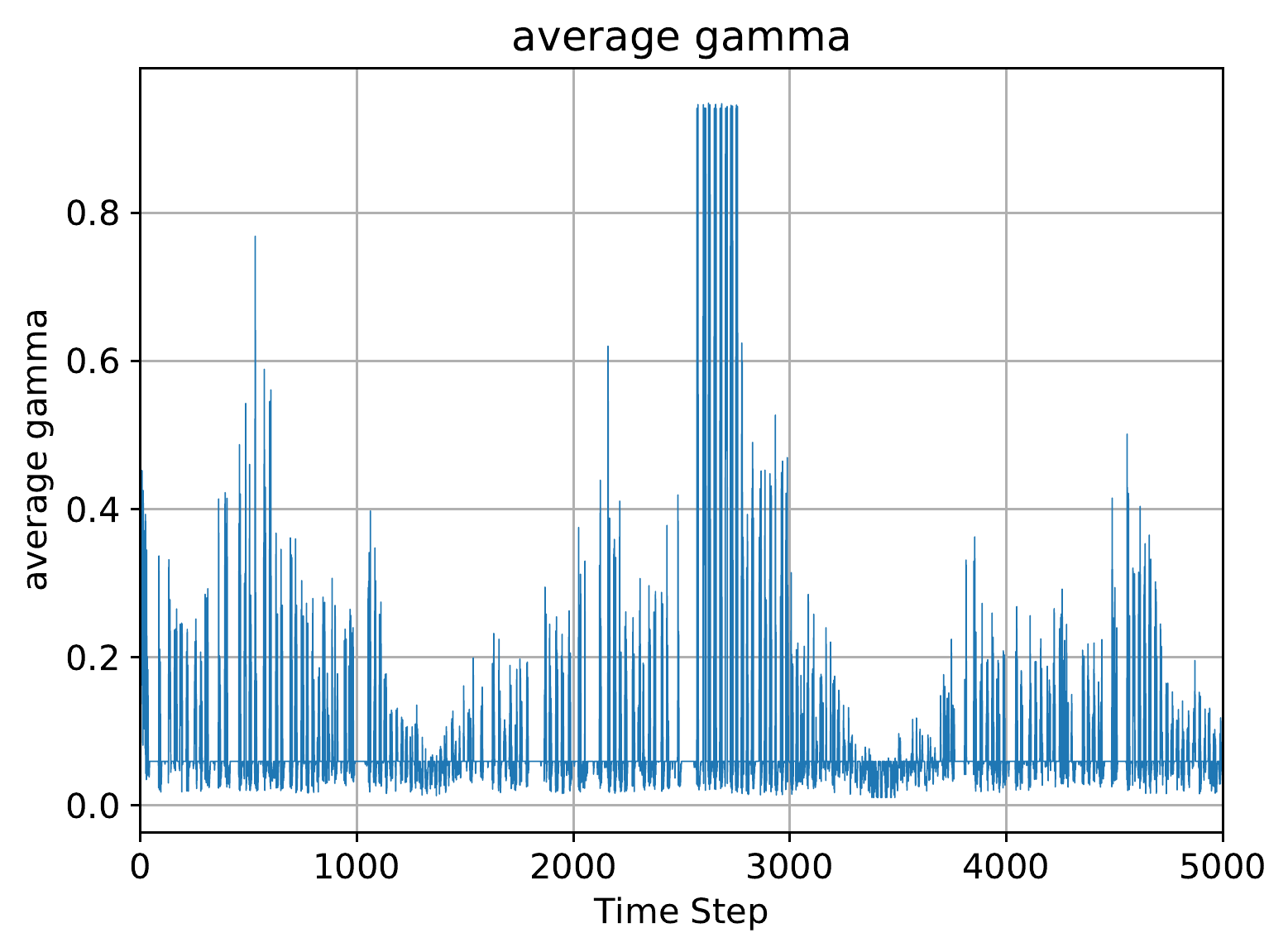}
\caption{$\xi$ = 0.75}
\label{fig:toleranz_vergleich_gamma_c}
\end{subfigure}
\begin{subfigure}{0.49\linewidth}
\includegraphics[width=\linewidth]{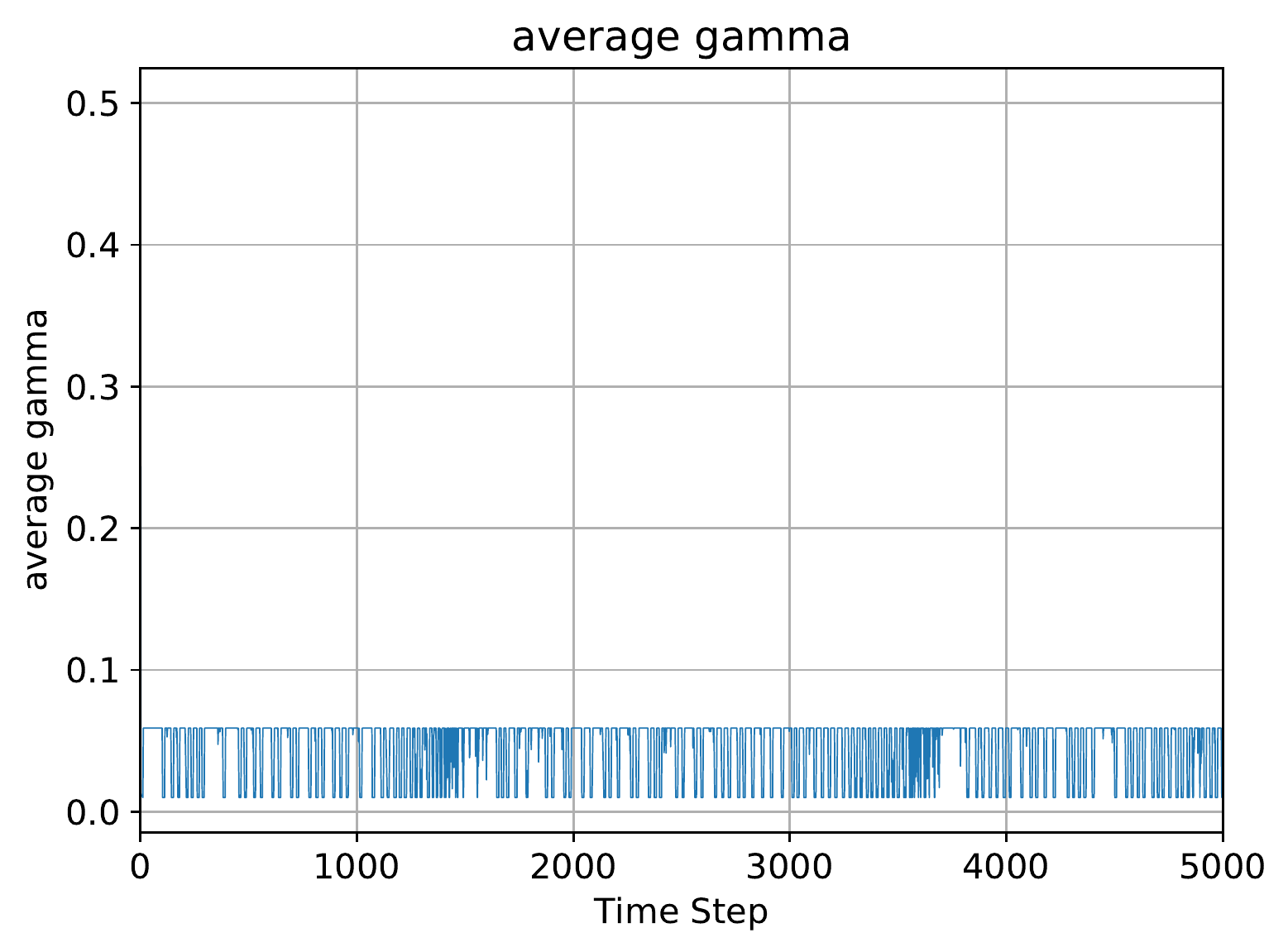}
\caption{$\xi$ = 1}
\label{fig:toleranz_vergleich_gamma_d}
\end{subfigure}
\caption{Average investment proportion for various tolerances.}
\label{fig:toleranz_vergleich_gamma}
\end{figure}

\begin{figure}[ht]
    \centering
    \begin{subfigure}{0.48\linewidth}
        \includegraphics[width=\linewidth]{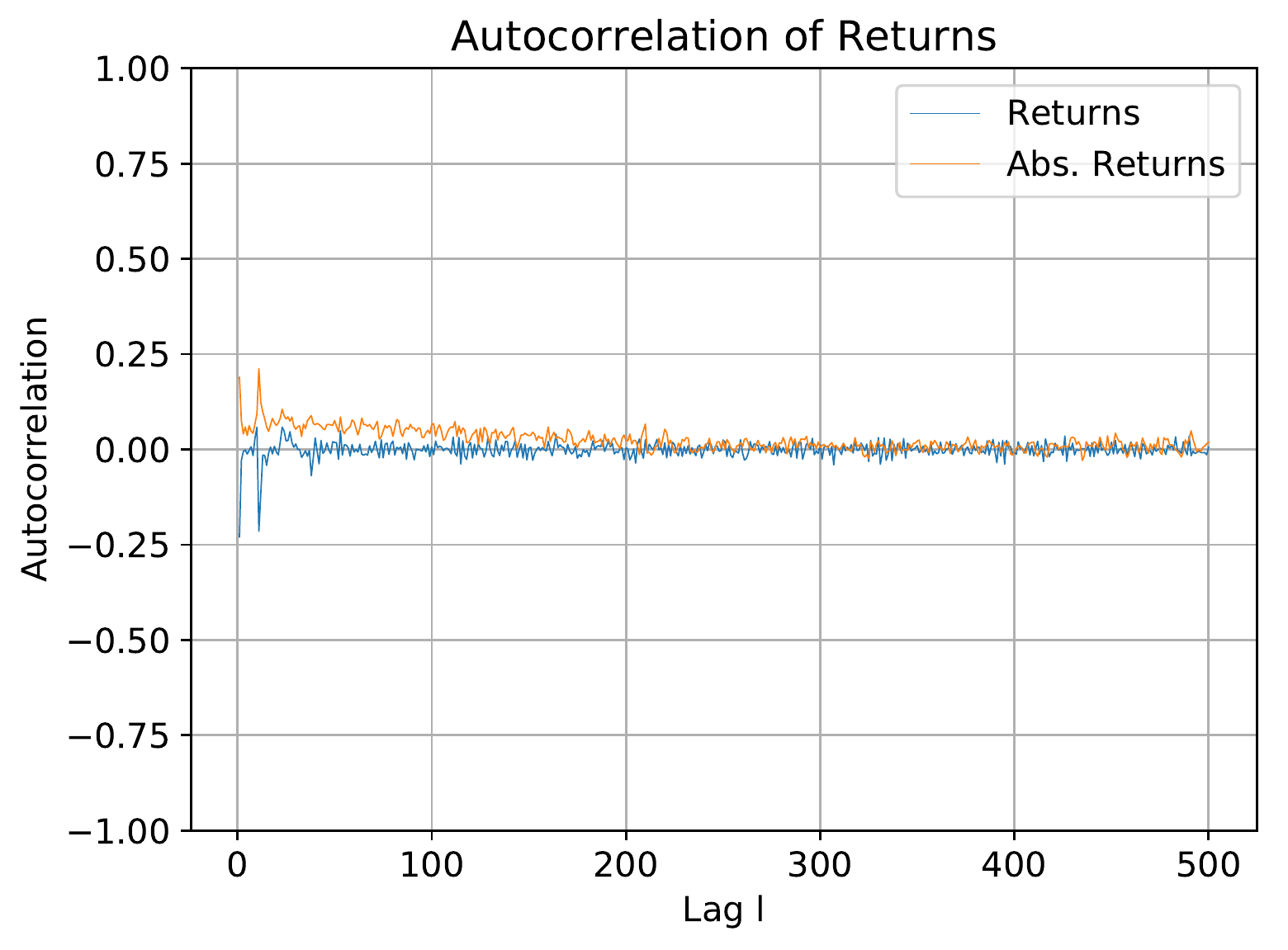}
        \caption{$\xi$ = 0.05}
        \label{fig:lls_autocorr_tolerances_a}
    \end{subfigure}
    \begin{subfigure}{0.48\linewidth}
        \includegraphics[width=\linewidth]{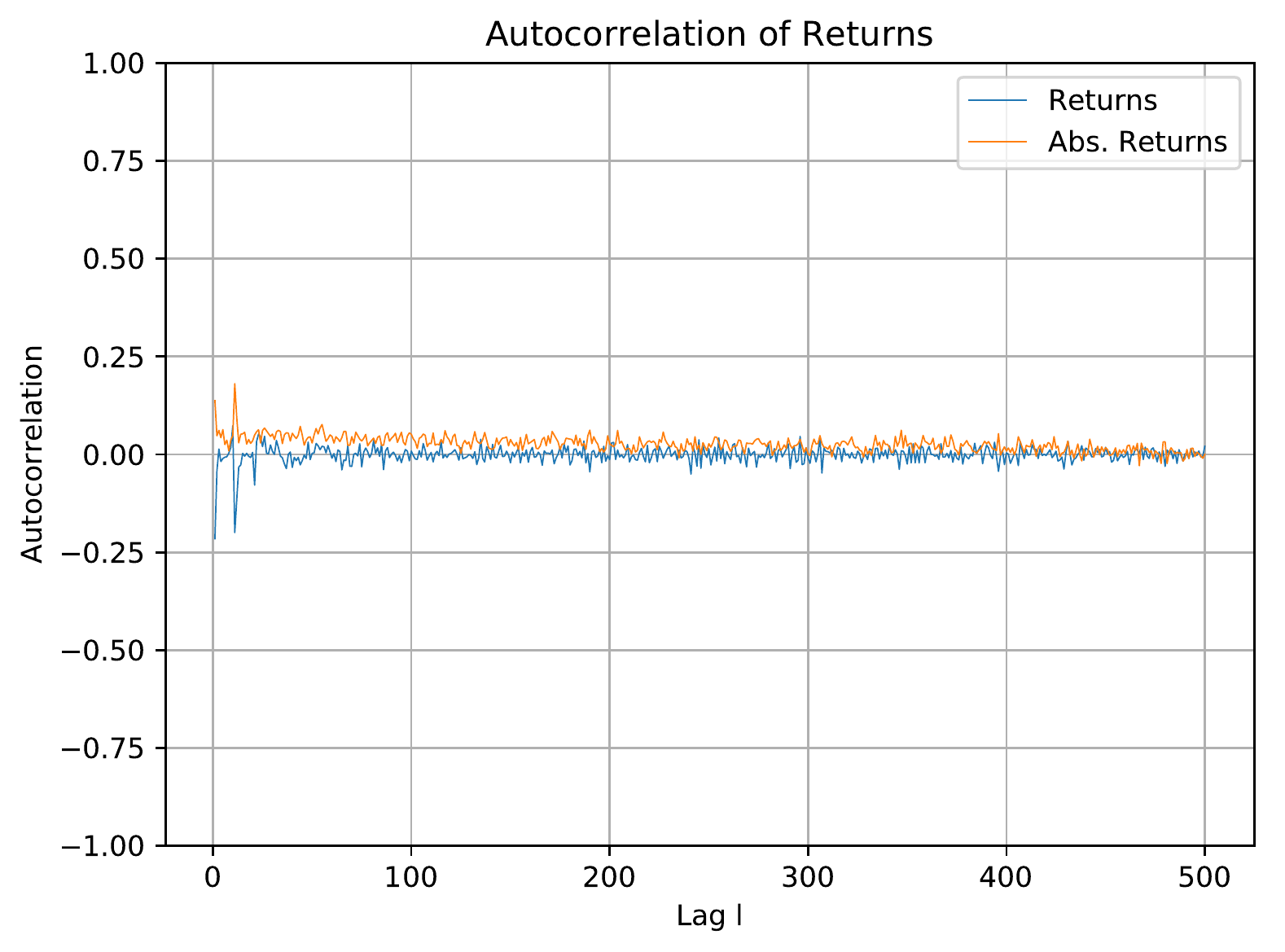}
        \caption{$\xi$ = 0.1}
        \label{fig:lls_autocorr_tolerances_b}
    \end{subfigure}
    \begin{subfigure}{0.48\linewidth}
        \includegraphics[width=\linewidth]{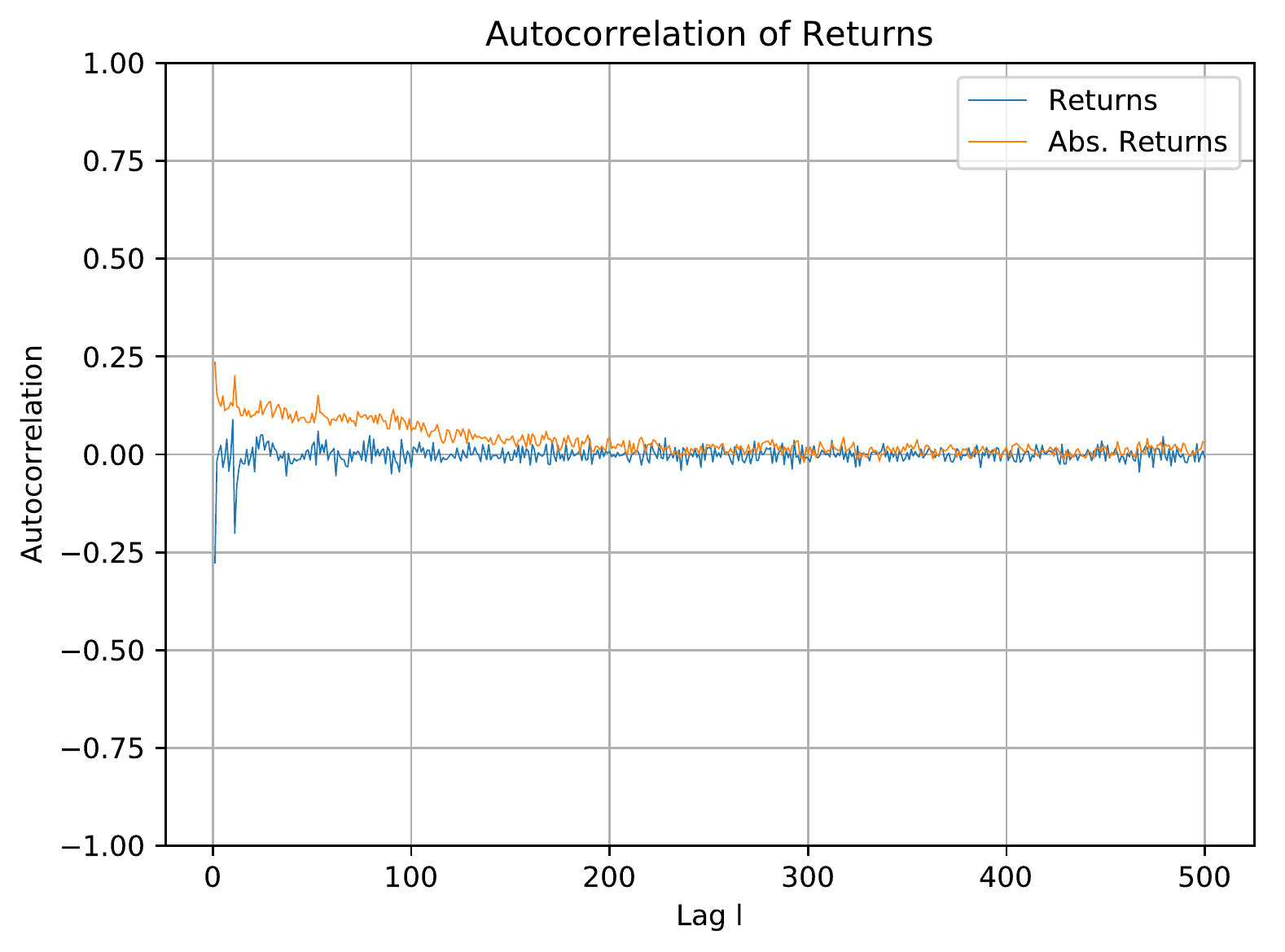}
        \caption{$\xi$ = 0.25}
        \label{fig:lls_autocorr_tolerances_c}
    \end{subfigure}
    \begin{subfigure}{0.48\linewidth}
        \includegraphics[width=\linewidth]{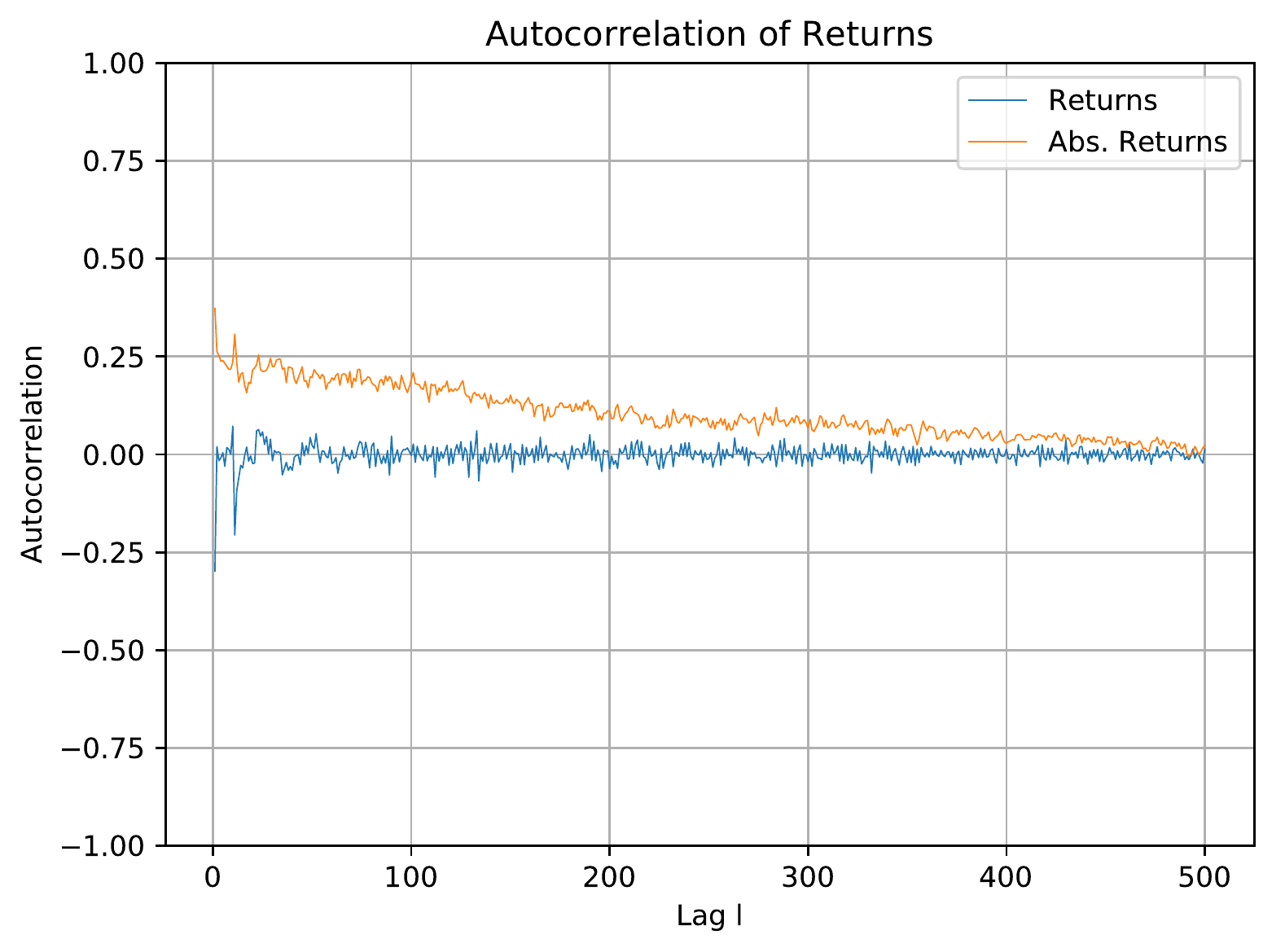}
        \caption{$\xi$ = 0.5}
        \label{fig:lls_autocorr_tolerances_d}
    \end{subfigure}
    \begin{subfigure}{0.48\linewidth}
        \includegraphics[width=\linewidth]{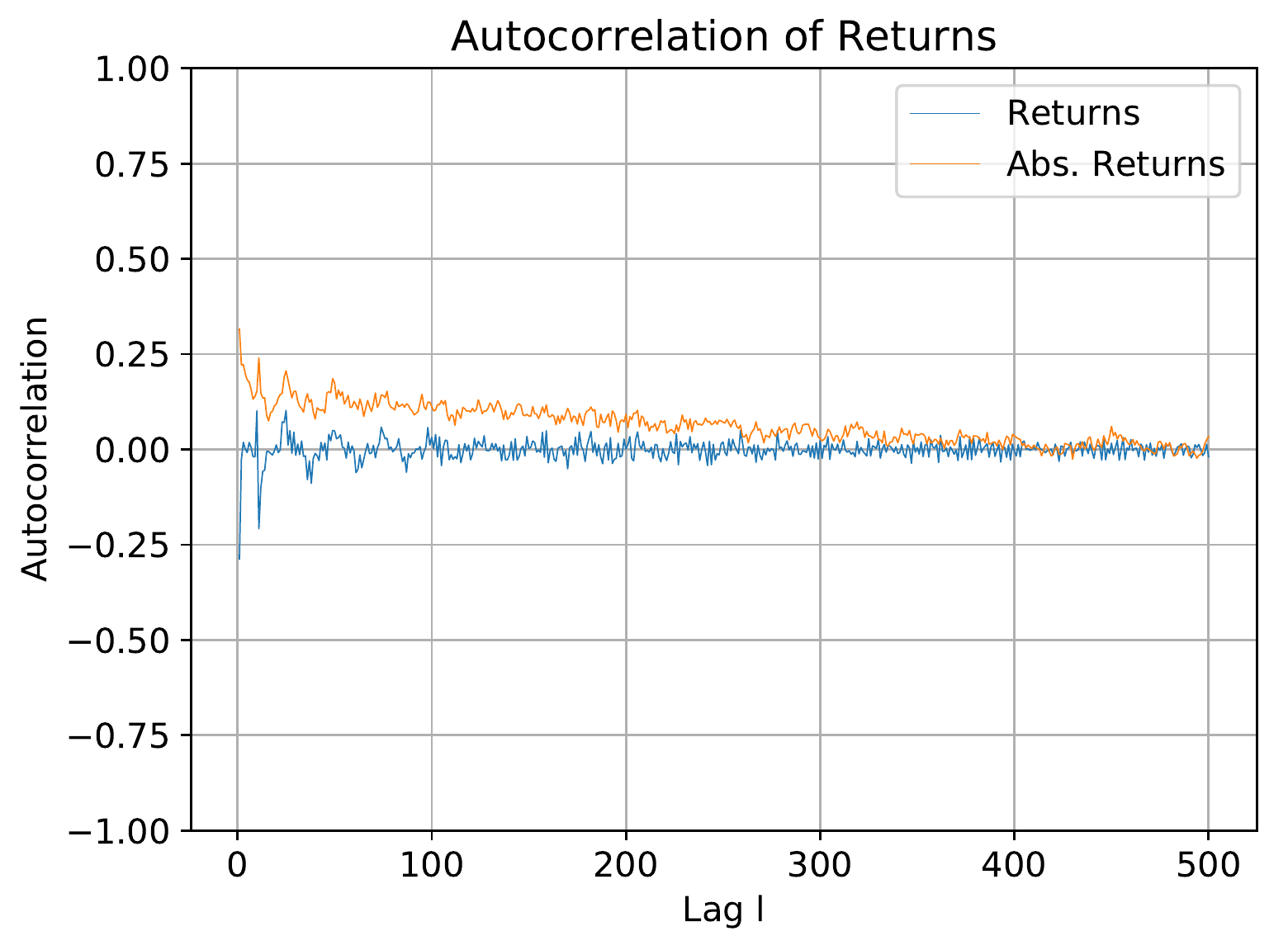}
        \caption{$\xi$ = 0.75}
        \label{fig:lls_autocorr_tolerances_e}
    \end{subfigure}
    \begin{subfigure}{0.48\linewidth}
        \includegraphics[width=\linewidth]{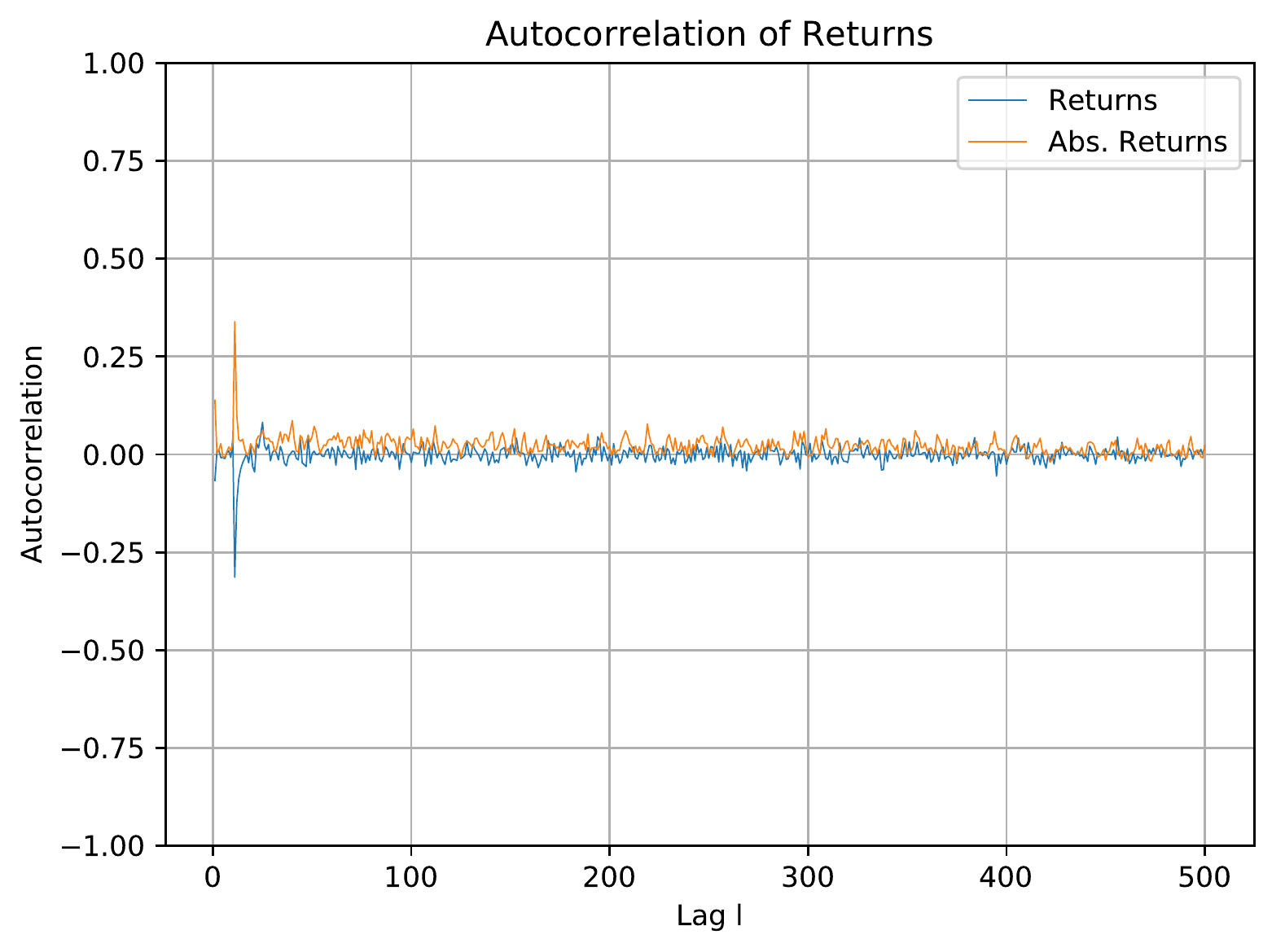}
        \caption{$\xi$ = 1}
        \label{fig:lls_autocorr_tolerances_f}
    \end{subfigure}
    \caption{Autocorrelation for various tolerances. For colored plots, please refer to online version.}
    \label{fig:lls_autocorr_tolerances}
\end{figure}

In this section, we study the impact of the stopping criterion applied in the root finding algorithm of the LLS model. This is employed in the clearance mechanism which fixes the stock price at each time step.
The clearance mechanism of the LLS model
$$n=\sum\limits_{k=1}^N \frac{\gamma_k(t)\ w_k(t)}{S(t)}$$
clearly models an equilibrium market where supply equals demand. In order to compute the equilibrium price at each time step, the clearance condition is solved numerically using a root finding algorithm. The crucial parameter is the chosen stopping criterion, defined as $\left| \sum\limits_{k=1}^N \frac{\gamma_k(t)\ w_k(t)}{S^k}-n \right|   \leq \xi,\ \xi>0$. Thus, the algorithm terminates if the clearance mechanism for the stock price $S^k$ is satisfied with respect to precision $\xi$.
 Note that the original publications \cite{levy1994microscopic, levy1995microscopic} do not report values or references for the tolerance used.\\
The choice of stopping criterion is also important from an economical view on the model. For relatively small values of $\xi$, the market price clearly is forced close to equilibrium resulting in an approximately rational market.
For larger $\xi$, the market may become irrational at times.  Note that the market is not inherently irrational for larger tolerances - most of the time, we observe low excess demand despite high tolerances. It is only around crashes that we find high excess demands that actually exhausts the tolerance. From a computational perspective, it is desirable to increase $\xi$ as much as possible to lower computational cost.  Figures \ref{fig:toleranz_vergleich_prices}, \ref{fig:toleranz_vergleich_gamma}, \ref{fig:lls_autocorr_tolerances} (200 agents) demonstrate that the choice of stopping criterion $\xi$ may strongly impacts the model behavior.

\begin{figure}[ht]
\centering
\begin{subfigure}{0.49\linewidth}
\includegraphics[width=\linewidth]{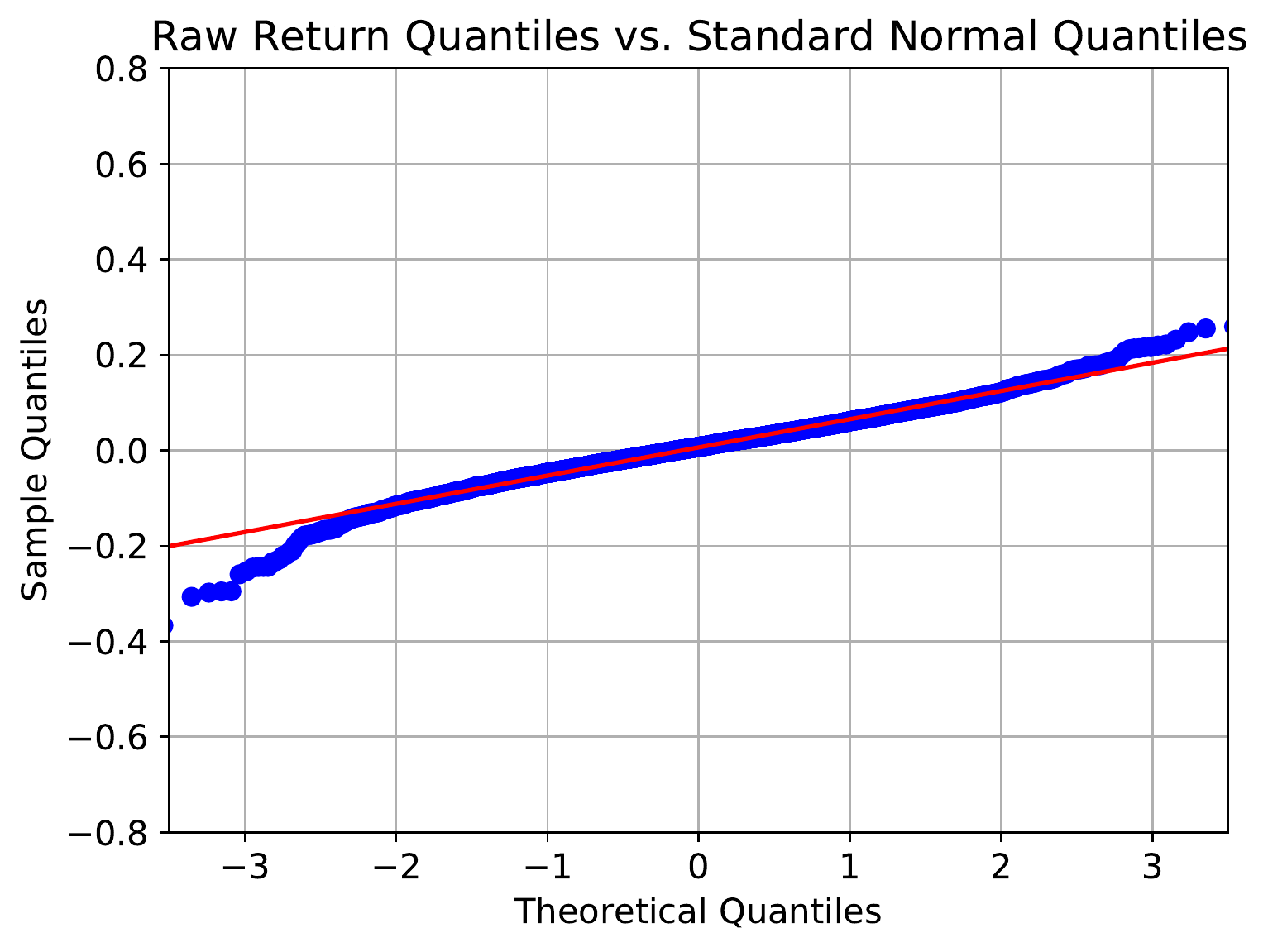}
\caption{$\xi$ = 0.1}
\label{fig:toleranz_qq_a}
\end{subfigure}
\begin{subfigure}{0.49\linewidth}
\includegraphics[width=\linewidth]{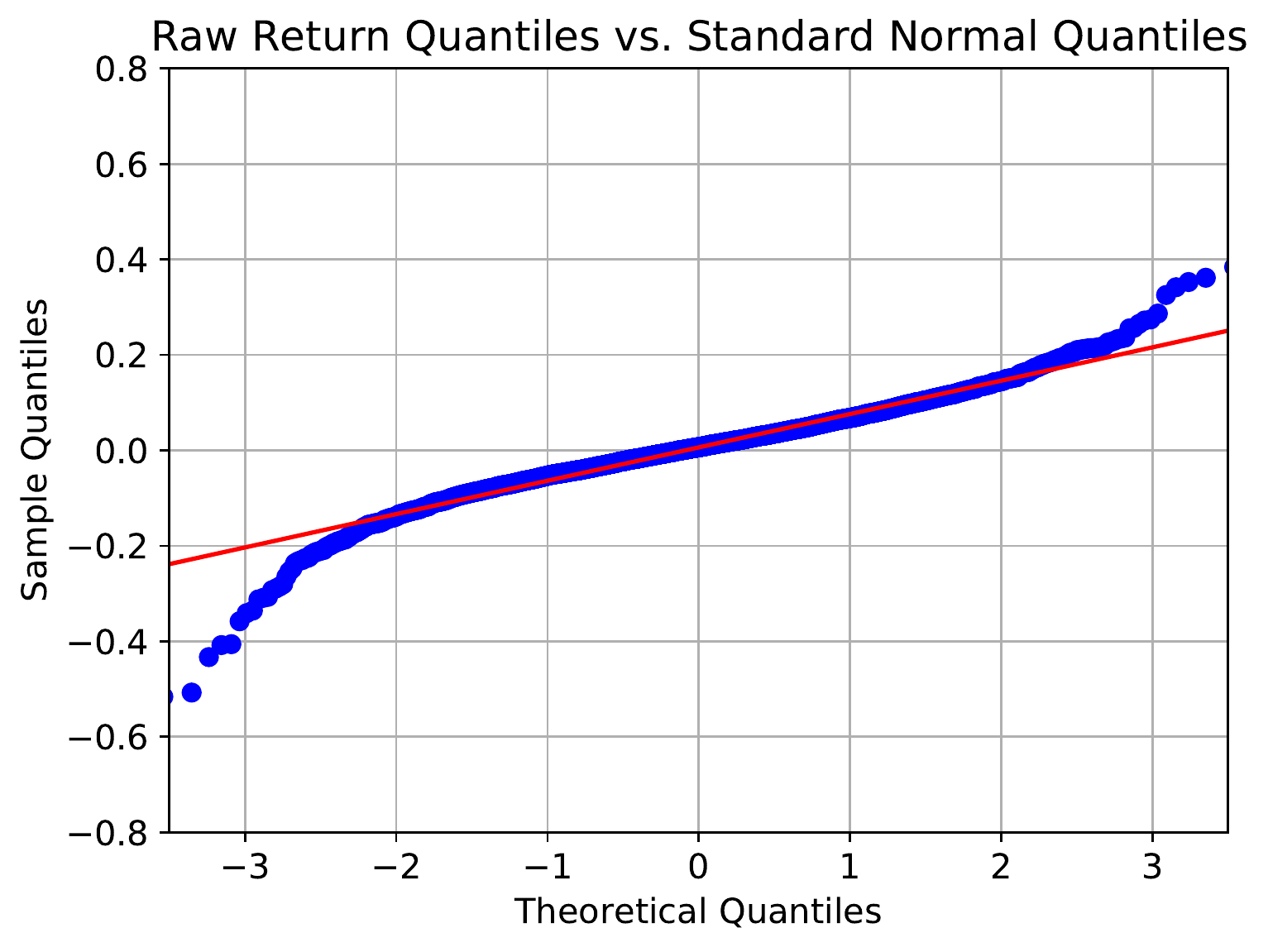}
\caption{$\xi$ = 0.25}
\label{fig:toleranz_qq_b}
\end{subfigure}
\begin{subfigure}{0.49\linewidth}
\includegraphics[width=\linewidth]{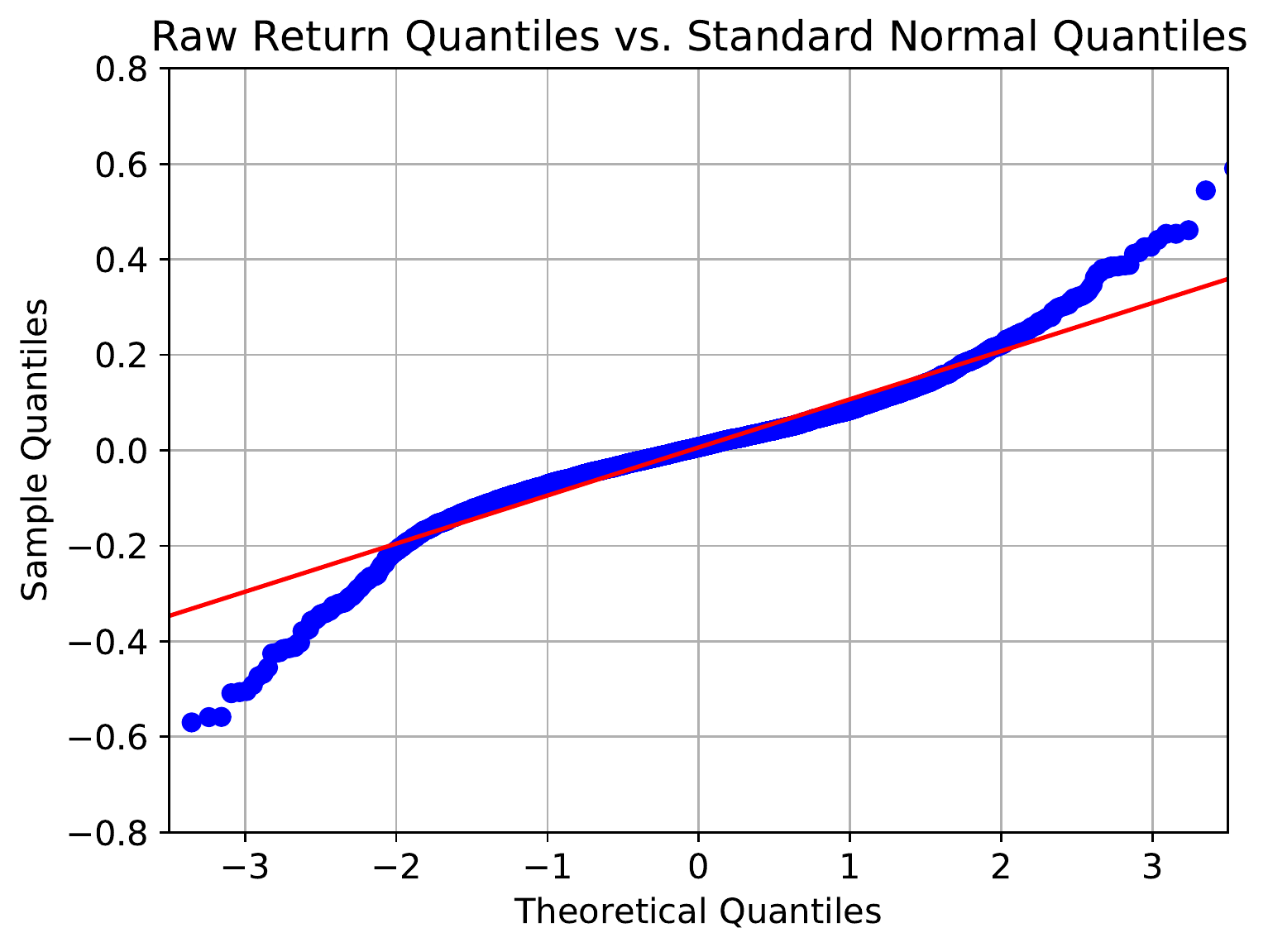}
\caption{$\xi$ = 0.5}
\label{fig:toleranz_qq_c}
\end{subfigure}
\begin{subfigure}{0.49\linewidth}
\includegraphics[width=\linewidth]{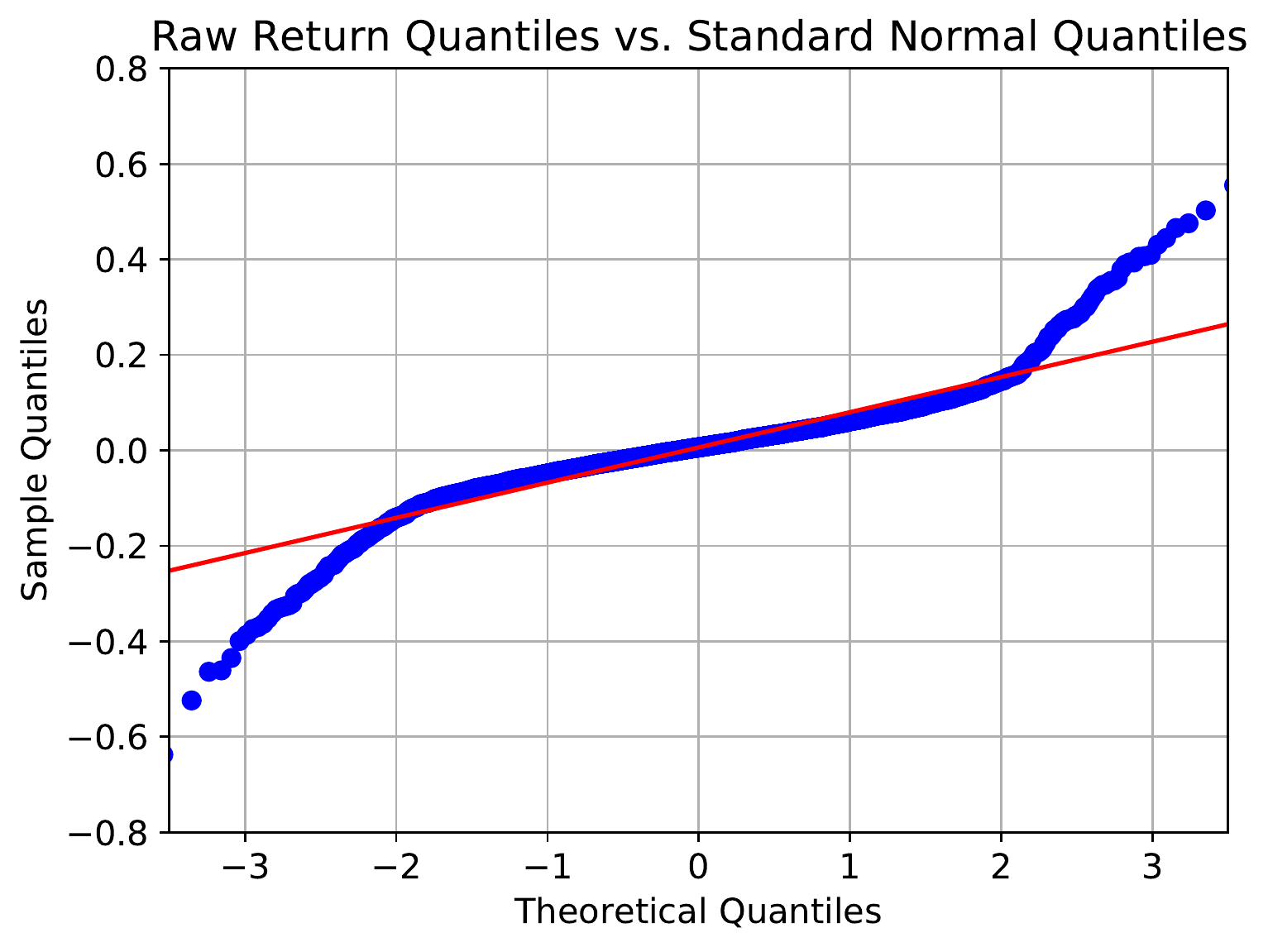}
\caption{$\xi$ = 1}
\label{fig:toleranz_qq_d}
\end{subfigure}
\caption{Qantile-quantile plot of returns for various tolerances.}
\label{fig:toleranz_qq}
\end{figure}

In Figure \ref{fig:toleranz_vergleich_prices}, we present results with tolerance levels $\xi=0.05$ and $\xi=0.1$. There are only subtle differences in return distribution and price trend.  In Figure \ref{fig:toleranz_vergleich_gamma}, simulation results for tolerances $\xi=0.1$, $\xi=0.5$, $\xi=0.75$, and $\xi=1$ are compared with respect to average investment proportions. There are moderate differences in the plots for $\xi \leq 0.75$. Notably, these results feature peaks at the same times and the trends look generally alike. For $\xi = 1$ however, average investment proportion is overall very low and it never exceeds 0.1.  Hence, $\xi = 1$ is found to represent too much of a relaxation of the rational market assumption.
 Figure \ref{fig:lls_autocorr_tolerances} presents the autocorrelation of logarithmic stock returns and absolute logarithmic stock returns for multiple tolerance levels.
Notably, the highest autocorrelation can be observed for $\xi = 0.25$, $\xi = 0.5$, and $\xi = 0.75$.
From qq-plots of logarithmic stock returns (see Figure \ref{fig:toleranz_qq}) one can conclude that increased tolerances result in heavier tails.
From an economic point of view, a non-negligible tolerance represents a relaxation of the rational market hypothesis and therefore results in an irrational market.
This raises the question whether increasing tolerance, and by this more irrational markets, generally result in heavier tails.

\section{Conclusion}
\label{sec-conclusion}
We have introduced the Levy-Levy-Solomon model and have conducted several numerical tests. 
As documented previously in \cite{zschischang2001some, kohl1997influence}, we have shown that  the Levy-Levy-Solomon model exhibits finite-size effects. 
This work indicates that finite-size effects are caused by a wrong scaling which leads to a variance reduction of the investment decision for large number of agents. 
In addition, we have verified that a high quality pseudo random number generator is essential to guarantee correct simulation results. This is of major importance in ABCEM models since many pseudo random numbers are needed and they are sensitive with respect to different noise levels as e.g. the Levy-Levy-Solomon model. 
Furthermore we studied the impact of different stopping criteria of the clearance mechanism on the model behavior. We have seen that the auto-correlation function of absolute returns and the heaviness of the fat-tails are influenced by the stopping criteria. Especially it seems that a higher tolerance leads to heavier tails, which maybe translated economically as follows: The more irrational the market is, the heavier are the tails of the stock return distribution of the Levy-Levy-Solomon model. This observation is in agreement with economic hypotheses on the influence of irrational markets on stock price behavior.
Clearly this finding deserves a more detailed study which is left open for future research.

\section*{Acknowledgement}
T. Trimborn was funded by the Deutsche Forschungsgemeinschaft (DFG, German Research Foundation) under Germany's Excellence Strategy – EXC-2023 Internet of Production – 390621612.
T. Trimborn gratefully acknowledges support by the Hans-Böckler-Stiftung and the RWTH Aachen University Start-Up grant. 
T. Trimborn acknowledges the support by the ERS Prep Fund - Simulation and Data Science. 
The work was partially funded by the Excellence Initiative of the German federal and state governments.

\appendix

\section{Appendix}
\label{sec-appendix}

\subsection{Parameter sets}\label{parameter}

\paragraph{LLS Model}
The initialization of the stock return is performed by creating an artificial history of stock returns. The artificial history is modeled as a Gaussian random variable with mean
$\mu_h$ and standard deviation $\sigma_h$. Furthermore, we have to point out that the increments of the dividend is deterministic, if $z_1=z_2$ holds. We used the C++  standard random number generator for all simulations of the LLS model if not otherwise stated. 
\begin{table}
	\begin{subtable}[b]{0.45\textwidth}
		\begin{center}
			\begin{tabular}{|c||c|}
			\hline
			Parameter & Value\\
			\hline
			\hline
			$N$ & $100$\\
			\hline
			$m_i$& $15$\\
			\hline 
			$\sigma_{\gamma}$ & $ 0 $ or $0.2$\\
			\hline
			$r$& $0.04$ \\
			\hline 
			$z_1=z_2$& $0.05$\\
			\hline
			time steps & 200 \\
			\hline
			\end{tabular}
		\end{center}
		\caption{Parameters of LLS model.}
	\end{subtable}
	\hspace{0.5cm}
	\begin{subtable}[b]{0.45\textwidth}
		\begin{center}
			\begin{tabular}{|c||c|}
			\hline
			 Variable & Initial Value\\
			\hline
			\hline
			$\mu_h$ & $0.0415$\\
			\hline
			$\sigma_h$ & $0.003$\\
			\hline
			$\gamma(t=0)$ & $0.4$\\
			\hline
			 $w_i(t=0)$ & $1000$ \\
			 \hline
			 $n_i (t=0)$ & $100$ \\
			 \hline
			 $S (t=0)$ & $4$ \\
			 \hline
			 $D (t=0)$ & $0.2$ \\
			 \hline
			\end{tabular}
		\end{center}
		\caption{Initial values of LLS model.}
	\end{subtable}
	\caption{Basic setting of the LLS model.} \label{LLS-basic}
\end{table}

\begin{table}
	\begin{subtable}[b]{0.45\textwidth}
		\begin{center}
			\begin{tabular}{|c||c|}
			\hline
			Parameter & Value\\
			\hline
			\hline
			$N$ & $99$\\
			\hline
			$m_i $&  $10,\ 1 \leqslant i \leqslant 33$  \\
			                  & $141,\ 34 \leqslant i \leqslant 66$ \\
			                  & $256,\ 67 \leqslant i \leqslant 99$ \\
			\hline 
			$\sigma_{\gamma}$ & $ 0.2 $\\
			\hline
			$r$& $0.0001$ \\
			\hline 
			$z_1=z_2$& $0.00015$\\
			\hline
			time steps & $20,000$\\
			\hline
			\end{tabular}
		\end{center}
		\caption{Parameters of LLS model.} 
	\end{subtable}
	\hfill
	\begin{subtable}[b]{0.45\textwidth}
		\begin{center}
			\begin{tabular}{|c||c|}
			\hline
			 Variable & Initial Value\\
			\hline
			\hline
			  $\mu_h$ & $0.0415$ \\
			  \hline
			  $\sigma_h$ &$ 0.003$ \\
			  \hline
			  $\gamma_i(t=0)$ & $0.4$ \\
			  \hline
			  $w_i (t=0)$ & $1000$ \\
			  \hline
			   $n_i (t=0)$ & $100$ \\
			   \hline
			    $S (t=0)$ & $4 $\\
			    \hline
			   $D (t=0)$ & $0.004$ \\
			 \hline
			\end{tabular}
		\end{center}
		\caption{Initial values of LLS model.}
	\end{subtable}
	\caption{Setting for the LLS model (3 agent groups).}
	\label{LLS-3-agents}
\end{table}

\begin{table}
    \centering
    \begin{tabular}{|c||c|}
        \hline
        Simulation & Random Number Generator \\
        \hline
        \hline
        \Crefrange{fig:embfinitesize_wealth_a}{fig:embfinitesize_wealth_d} & C++ MT19937 RNG ($64$ bit)\\
        \hline
        \Crefrange{fig:emb_fse_erklarung_a}{fig:emb_fse_erklarung_c} & C++ MT19937 RNG ($64$ bit)\\	            
        \hline
        \Cref{qual-rng_a} & C++ MT19937 RNG ($64$ bit)\\	            
        \hline
        \Cref{qual-rng_b} & \texttt{RANDU} generator ($32$ bit)\\	                 
        \hline
                 \Crefrange{fig:toleranz_vergleich_prices_a}{fig:toleranz_vergleich_prices_b} & C++ MT19937 RNG ($64$ bit)\\	            
        \hline
                 \Crefrange{fig:toleranz_vergleich_gamma_a}{fig:toleranz_vergleich_gamma_d} & C++ MT19937 RNG ($64$ bit)\\	            
        \hline
                 \Crefrange{fig:toleranz_vergleich_gamma_a}{fig:toleranz_vergleich_gamma_d} & C++ MT19937 RNG ($64$ bit)\\            
        \hline

                         \Crefrange{fig:lls_autocorr_tolerances_a}{fig:lls_autocorr_tolerances_f} & C++ MT19937 RNG ($64$ bit)\\            
        \hline
                         \Crefrange{fig:toleranz_qq_a}{fig:toleranz_qq_d} & C++ MT19937 RNG ($64$ bit)\\                  
        \hline
    \end{tabular}
    \caption{Random Generators for the Simulations}
    \label{tab:RNGTable}
\end{table}

%-- LITERATUR ----------------------------------------------------------%
	%\clearpage
	%\addcontentsline{toc}{section}{Literaturverzeichnis}
	\bibliography{SABCEMM.bib}
	\bibliographystyle{abbrvdin}

\end{document}